\documentclass[twocolumn]{aastex7}

\usepackage{graphicx}   
\usepackage{textcomp}   
\usepackage{amsmath}
\usepackage{lineno}
\usepackage{subcaption}
\usepackage{float}
\usepackage{afterpage}
\usepackage{booktabs}
\usepackage{siunitx}
\usepackage[table]{xcolor}
\usepackage{multirow} 
\usepackage{array}

\graphicspath{{./}{figures/}}
\shorttitle{galactic outflows at cosmic noon}
\shortauthors{Chen, Zhu, Li, Wang et al.}

\begin{document}
\title{Properties of Galactic Outflows Driven by Starburst at Cosmic Noon: Insights from Hydrodynamical Simulations}

\author{Huan Chen}
\email{chenh699@mail2.sysu.edu.cn}
\affil{School of Physics and Astronomy, Sun Yat-Sen University, Zhuhai campus, No. 2, Daxue Road \\
Zhuhai, Guangdong, 519082, China}
\affil{CSST Science Center for the Guangdong-Hong Kong-Macau Greater Bay Area, Daxue Road 2, 519082, Zhuhai, China}

\author[0000-0002-1189-2855]{Weishan Zhu}
\email{zhuwshan5@mail.sysu.edu.cn}
\affil{School of Physics and Astronomy, Sun Yat-Sen University, Zhuhai campus, No. 2, Daxue Road \\
Zhuhai, Guangdong, 519082, China}
\affil{CSST Science Center for the Guangdong-Hong Kong-Macau Greater Bay Area, Daxue Road 2, 519082, Zhuhai, China}

\author{Xue-Fu Li}
\email{lixf96@mail2.sysu.edu.cn}
\affil{School of Physics and Astronomy, Sun Yat-Sen University, Zhuhai campus, No. 2, Daxue Road \\
Zhuhai, Guangdong, 519082, China}
\affil{CSST Science Center for the Guangdong-Hong Kong-Macau Greater Bay Area, Daxue Road 2, 519082, Zhuhai, China}

\author{Tian-Rui Wang}
\email{wangtr3@mail2.sysu.edu.cn}
\affil{School of Physics and Astronomy, Sun Yat-Sen University, Zhuhai campus, No. 2, Daxue Road \\
Zhuhai, Guangdong, 519082, China}
\affil{CSST Science Center for the Guangdong-Hong Kong-Macau Greater Bay Area, Daxue Road 2, 519082, Zhuhai, China}

\author{Antonios Katsianis}
\email{katsianis@mail.sysu.edu.cn}
\affil{School of Physics and Astronomy, Sun Yat-Sen University, Zhuhai campus, No. 2, Daxue Road \\
Zhuhai, Guangdong, 519082, China}
\affil{CSST Science Center for the Guangdong-Hong Kong-Macau Greater Bay Area, Daxue Road 2, 519082, Zhuhai, China}

\author{Long-Long Feng}
\email{flonglong@mail.sysu.edu.cn}
\affil{School of Physics and Astronomy, Sun Yat-Sen University, Zhuhai campus, No. 2, Daxue Road \\
Zhuhai, Guangdong, 519082, China}
\affil{CSST Science Center for the Guangdong-Hong Kong-Macau Greater Bay Area, Daxue Road 2, 519082, Zhuhai, China}

\correspondingauthor{Weishan Zhu}
\email{zhuwshan5@mail.sysu.edu.cn}



\begin{abstract}
We investigate starburst-driven galactic outflows in low-mass galaxies ($9.0 < \log(M_*/M_\odot) < 10.0$) at cosmic noon using high-resolution 3D hydrodynamical simulations based on a framework that can reproduce the multiphase outflows in M82. The simulations produce starbursts lasting 20-30 Myr, with peak star formation rates of 2-68 M$_\odot \,\rm{yr}^{-1}$. Outflow properties vary strongly with time, radial distance to galaxy center, stellar mass, and gas fraction, exhibiting velocities of 50-1000 $\,\rm{km\,s}^{-1}$, mass outflow rates of 0.3-20 M$_\odot \,\rm{yr}^{-1}$, and mass loading factors, $\eta_\mathrm{M}$, of 0.24-6.26. The cool phase ($8000 < T \le 2 \times 10^4$ K) dominates the outflow, and properties of the cool and warm phases are broadly consistent with observations. At $M_*= 10^{9.5}\,M_\odot$, average $\eta_\mathrm{M}$ for the total, cool, and warm phases are $\sim$1.2, 0.75, and 0.25, respectively. The mass loading factor decreases with increasing galaxy stellar mass, but increases with star formation rate. Given strong temporal and spatial evolution, scaling slopes from limited samples should be treated with caution. Our total $\eta_\mathrm{M}$ values are higher than FIRE-2 by 0.06 dex but lower than EAGLE and TNG50 by 0.50 and 0.84 dex. Accounting for methodological differences in outflow measurement reduces these gaps to 0.2–0.4 dex, suggesting that part of the discrepancy between observations and simulations reported in the literature may arise from inconsistent definitions and measurement methods, though differences in individual phases persist. Larger observational and simulation samples, together with consistent methods for measuring outflow properties, are required to draw robust conclusions about the scaling relations of galactic outflows.

\end{abstract}




\section{Introduction} \label{sec:intro}
Galactic outflows have been observed in both high-redshift galaxies \citep[e.g.,][]{2003ApJ...588...65S, schreiber2019kmos3d, swinbank2019energetics, 2022MNRAS.515..841W, concas2022being, 2023ApJ...949....9P, 2024MNRAS.531.4560W, cooper2025high} and galaxies in the local Universe \citep[e.g.,][]{1963ApJ...137.1005L, 1999ApJ...513..156M, 2005ApJS..160..115R, 2014A&A...568A..14A, 2019ApJ...886...74M, 2022ApJ...933..222X}. Driven by either stellar or AGN feedback, galactic outflows, also known as galactic winds, are crucial components of the galaxy ecosystem. They play a vital role in galaxy evolution by expelling gas from the host galaxy, thereby regulating its baryon content, controlling star formation efficiency, and enriching the circumgalactic medium (CGM) and intergalactic medium (IGM) with metals \citep[e.g.,][]{1986ApJ...303...39D, 1994MNRAS.271..781C, 1999ApJ...519L.109C, 2003ApJ...599...38B, 2005ARA&A..43..769V, 2017MNRAS.465.1682H, 2017ARA&A..55...59N}. In this work we focus on galactic winds driven by starbursts in relatively low-mass galaxies at cosmic noon.

Over the past several decades, sustained efforts from observational and theoretical studies have contributed to a broad understanding of the launching and evolution of starburst-driven galactic winds \citep[e.g.,][]{1985Natur.317...44C,1988ApJ...330..695T,Tomisaka1993,Strickland2000,2005ApJ...618..569M,2005ARA&A..43..769V,2011ApJ...735...66M,2017arXiv170109062H,2018Galax...6..114Z,2024ARA&A..62..529T}. Over timescales of several to tens of Myr, supernovae release large amounts of momentum and energy into the interstellar medium (ISM) in the central regions of galaxies. A portion of this feedback energy is converted into thermal energy, heating the surrounding ISM into distinct temperature phases: cool ($8000 < T \le 2\times10^4$ K), warm ($2\times10^4 < T \le 5\times10^5$ K), and hot ($T > 5\times10^5$ K). This multiphase gas is then driven outward by stellar feedback at velocities ranging from several hundred to a couple of thousand $\,\rm{km\,s}^{-1}$, ultimately forming galactic outflows that can extend to scales of up to tens of kpc.

However, the physical mechanisms of  starburst driven galactic winds remain incompletely understood \citep[e.g.,][]{2017arXiv170109062H, 2018Galax...6..114Z, 2020A&ARv..28....2V, 2024ARA&A..62..529T}. Major uncertainties persist in the star formation history of individual starbursts, the spatial distribution of newly formed stars, and the fraction of core-collapse supernovae per massive star, leading to large uncertainties in the total injected feedback energy. Moreover, how stellar feedback, particularly supernovae, couples to the inhomogeneous ISM and transfers energy and momentum to multiphase outflows remains poorly constrained. Key wind properties, including the thermalization efficiency of hot phase, the partitioning between thermal and kinetic energy, and mass loading factors  vary widely across theoretical models \citep{1985Natur.317...44C, 2005ApJ...618..569M, 2022ApJ...924...82F}. The mixing and energy, momentum exchange among different wind phases, and their impact on outflow properties, also require further investigation.

Meanwhile, the origin of substantial amounts of dense cold phases observed in the galactic winds is not determined. Mechanisms responsible for accelerating cool gas to the high velocities inferred from observations are still an open question \citep[e.g.,][]{2020A&ARv..28....2V, 2024ARA&A..62..529T}. It is also unclear how galactic outflows redistribute metals within the wind and the circumgalactic medium across different spatial scales \citep[e.g.,][]{2023ARA&A..61..131F}. In addition, the roles of radiation pressure and cosmic rays in driving and shaping galactic winds require further investigation, owing to significant uncertainties in metallicity and gas-to-dust ratios, as well as in the microphysics governing cosmic-ray transport \citep[e.g.,][]{2008ApJ...687..202S, 2018ApJ...868..108B, 2022MNRAS.510..920Q, 2022MNRAS.510.1184Q}.

At the same time, the observed properties of galactic winds, such as velocities and mass outflow rates across different gas phases, are subject to substantial uncertainties \citep[e.g.,][]{2018Galax...6..138R}. Observations of outflows in low-mass galaxies are challenging due to the limited sensitivity of current ground-based near-infrared facilities, which hampers the detection of the faint broad components of nebular emission lines tracing ionized outflows \citep[e.g.,][]{2024MNRAS.531.4560W, concas2022being}. Moreover, the inferred outflow properties are highly sensitive to methodological choices, including emission-line fitting procedures and assumptions about outflow geometry (e.g., spherical versus biconical), temperature, and electron density.

Scaling relations between outflow properties (e.g., velocity, mass outflow rate, mass loading factor) and host galaxy characteristics (e.g., stellar mass, star formation rate) exhibit significant discrepancies across observational studies \citep[e.g.,][]{swinbank2019energetics, concas2022being, llerena2023ionized, 2024MNRAS.531.4560W, cooper2025high}.
For instance, some works have found that the mass loading factor of galactic outflows in galaxies with stellar masses of $10^{9}-10^{11} M_\odot$ remains nearly constant or shows a weak positive correlation with stellar mass \citep[e.g.,][]{2019ApJ...873..102F, schreiber2019kmos3d, swinbank2019energetics}. In contrast, more recent studies have reported the opposite trend, with the mass loading factor decreasing with increasing stellar mass \citep[e.g.,][]{concas2022being, 2024A&A...685A..99C, cooper2025high}.

Therefore, developing a more comprehensive understanding of starburst-driven galactic outflows is critical for interpreting multiphase wind properties and for refining theoretical models of galaxy formation and evolution. In particular, establishing robust relations between outflow characteristics and host galaxy properties, especially during cosmic noon, the peak epoch of cosmic star formation activity, remains a critical challenge. 

Hydrodynamical simulations are powerful tools for investigating the physical mechanisms driving galactic winds and their role in galaxy evolution. By incorporating galactic winds powered by stellar and AGN feedback through subgrid models, state-of-the-art cosmological hydrodynamical simulations, such as EAGLE and IllustrisTNG, have successfully reproduced many statistical properties of galaxies and generate galactic winds in star forming galaxies. However, their limited spatial and mass resolution precludes directly resolving the multiphase ISM and its coupling to feedback processes \citep[e.g.,][]{2018MNRAS.473.4077P, 2019MNRAS.486.2827D}. As a result, these simulations often struggle to capture the full launching and evolution of multiphase gas outflows, lacking realistic multiphase galactic wind structures and physically self-consistent mass loading process \citep{2017ARA&A..55...59N,2023ARA&A..61..131F}.

For example, EAGLE relies on stochastic thermal stellar feedback. As a result, a substantial fraction of the injected supernova energy is radiated away in dense gas, producing relatively weak and slow winds that nonetheless efficiently entrain surrounding material \citep{mitchell2020galactic}. In contrast, IllustrisTNG suppresses gas cooling below $10^4$ K, which leads to a deficit of cold gas phases in the outflow \citep{nelson2019first}. Moreover, the mass loading factors predicted by EAGLE and IllustrisTNG are moderately and substantially higher, respectively, than those inferred from observations \citep[e.g.,][]{nelson2019first, mitchell2020galactic, 2023arXiv231006614X, 2024MNRAS.531.4560W, 2024A&A...685A..99C, cooper2025high, 2025MNRAS.tmp.1579S}.

In contrast, zoom-in simulations such as FIRE-2 \citep{pandya2021characterizing}, AURIGA \citep{2016MNRAS.459..199G}, APOSTLE \citep{2016MNRAS.457.1931S}, and LYRA \citep{2021MNRAS.501.5597G} achieve substantially higher spatial resolution by focusing on individual galaxies or groups, enabling more direct investigations of stellar feedback and baryon cycling processes. Nevertheless, different zoom-in simulations exhibit diverse baryon cycle behaviors, reflecting variations in how feedback regulates gas inflows and outflows \citep{2022MNRAS.514.3113K}. For example, the ionized-phase mass loading factors in FIRE-2 are more consistent with current observational constraints, likely in part because FIRE-2 adopts a stricter definition of outflows, counting only gas that is truly unbound and capable of escaping the galaxy \citep{2024ApJ...966..129K}. However, with typical resolutions of order $\sim100$ pc, these simulations still cannot fully resolve key processes such as star formation, the coupling of feedback with the multiphase ISM, and interactions among different wind phases.

High-resolution simulations of starburst-driven winds in isolated galaxies or gas discs, typically spanning volumes from a few hundred parsecs to several kiloparsecs on a side, provide a crucial window into the detailed structure and evolution of multiphase galactic winds. Simulations of starburst-driven outflows in gas discs, such as TIGRESS \citep{kim2017three, 2018ApJ...853..173K, 2020ApJ...903L..34K}, SILCC \citep{2016MNRAS.456.3432G}, and QUOKKA \citep{2024MNRAS.52710095V, 2025MNRAS.539.1706V}, can resolve small scale star formation and capture, in detail, the interactions between stellar feedback and the surrounding ISM. For example, \citet{fielding2018clustered} showed that outflow strength is highly sensitive to the spatial distribution and clustering of supernovae. These simulations further show that the hot phase carries the majority of the wind’s energy, whereas the cold phase dominates its mass budget \citep{2020ApJ...903L..34K, 2025MNRAS.539.1706V}. 

Recent idealized, high-resolution simulations of galactic winds in isolated galaxies have achieved mass resolutions of $\Delta m \sim 1$–$10^{3}\,M_\odot$ \citep[e.g.,][]{2018MNRAS.478..302S, 2019MNRAS.483.3363H, 2022ApJ...924...82F, 2024A&A...691A.231D, 2024ApJ...960..100S}, spatial resolutions of $\Delta x \sim 1$–$100$ pc, and simulated volumes spanning several cubic kiloparsecs \citep[e.g.,][]{schneider2020physical, 2024arXiv241209452W, 2025ApJ...982...28L}. At these resolutions, such simulations can resolve key physical processes and provide detailed insights into the launching, structure, and evolution of multiphase outflow. 

High-resolution simulations offer a powerful approach to studying starburst-driven galactic winds in environments comparable to nearby galaxies, enabling direct comparisons with observations. M82 is an ideal case, having undergone a recent starburst and exhibiting a well-observed, multiphase outflow across multiple wavelengths \citep[e.g.,][]{1985Natur.317...44C, 2005ApJ...618..569M, 2008ApJ...674..157C, schneider2020physical}. Numerous three-dimensional simulations of M82-like systems have reproduced many observed wind features, highlighting the importance of ISM structure, stellar cluster distribution, and multiphase interactions in shaping outflow properties (e.g., \citealt{2008ApJ...674..157C, 2008ApJ...689..153R, melioli2013evolution, schneider2020physical, schneider2024cgols}).

More recently, \citet{2024arXiv241209452W} and \citet{2025ApJ...982...28L} developed self-consistent simulations that reproduce key observed properties of M82's starburst and multiphase outflows, including starburst duration and intensity, wind morphology, mass outflow rates, and X-ray emission. These studies demonstrate that interactions between supernova feedback and a multiphase, inhomogeneous ISM, along with clustered supernovae and GMC disruption, plays a crucial role in shaping galactic winds.

To further test the robustness of this framework, we now extend it to isolated, star-forming, low-mass galaxies at cosmic noon, which allow us to solidify our finding about mechanisms of starburst-driven winds, and explore how wind properties depend on galaxy characteristics at $z \sim 1$–$2$. This paper is structured as follows. In Section~\ref{sec:method}, we describe our simulation setup, including the physical and galaxy models. Section~\ref{sec:results} presents the properties of the simulated starbursts and resulting outflows. In Section~\ref{sec:discuss}, we compare our results with existing studies and discuss caveats and directions for future work. Our main conclusions are summarized in Section~\ref{sec:conclusions}.

\section{Methodology} \label{sec:method}
We perform simulations of star formation and outflows in 14 idealized isolate low mass disk galaxies, with gas fractions ranging from $30\%$ to $85\%$, typical of $1<z<2$ galaxies (\citealt{daddi2010very, 2012MNRAS.426.1178N, 2019ApJ...878...83W}). Each simulation begins with specific initial conditions for the gas disc, stellar bulge and disc, and dark matter (DM) halo. We then follow the evolution of star formation and the development of galactic winds over a period of 30 Myr, using the framework established by \citet{2024arXiv241209452W} and \citet{2025ApJ...982...28L}. Section~\ref{subsec:Physics Modules} provides a brief overview of the baryonic physics included in our simulations, such as radiative cooling, star formation, and stellar feedback. Section~\ref{subsec:Galaxy Models} describes the galaxy models, which are designed to represent typical isolated disc galaxies at $z \sim 1$–$2$, but without cosmological environmental effects such as accretion, interactions and mergers.

\subsection{Physical Modules}\label{subsec:Physics Modules}
In our simulations, we initialize a gas disk and an isothermal hot gas halo within a cubic volume and evolve the system using the Athena++ code \citep{stone2020athena++}. The gravitational potentials of the DM halo, stellar disk, and bulge are modeled as static background fields, following the literature (e.g. \citealt{2008ApJ...674..157C,2018ApJ...853..173K} ). This approximation is reasonable, as the outflows in our simulations have only minor effect on the global potential. The self-gravity of the gas is computed using Athena++’s built-in gravity solver. Star formation, stellar feedback, and radiative cooling/heating are implemented through additional physics modules introduced in \citet{2024arXiv241209452W} and \citet{2025ApJ...982...28L}. 
At each timestep, the gravitational effects of newly formed stars and their interactions with the gas are fully accounted for. The initial gas metallicity in all simulations is set to 0.02 $Z_\odot$, and the evolution of metals is tracked using passive scalar fields. In \citet{2025ApJ...982...28L}, we find only minor differences in the outflow properties between simulations with initial metallicities of 0.02 $Z_\odot$ and 0.1 $Z_\odot$. Here, we adopt the lower value to better represent the gas conditions prior to starburst enrichment. A passive scalar is used to trace the evolution of metals in the ISM, accounting for enrichment from stellar winds and supernova ejecta.

Metallicity-dependent radiative cooling and heating are computed using the Grackle library (v3.3.dev1; \citealt{smith2017grackle})\footnote{https://grackle.readthedocs.io}
. Following \citet{2024arXiv241209452W}, we adopt Grackle’s tabulated cooling functions calculated with CLOUDY, which include collisional ionization, photoionization by the cosmic ultraviolet background (UVB; \citealt{2012ApJ...746..125H}), and self-shielding of HI and HeI from the UVB. In addition, we include photoelectric heating from newly formed stars in the cooling/heating source term, rather than explicitly modeling photoionization, for the reasons discussed in \citet{2024arXiv241209452W}. At each time step, the gas density, velocity, temperature, metallicity, and local UV flux, computed as described in \citet{2024arXiv241209452W}, are passed to Grackle to evaluate the corresponding cooling and heating rates.

As described in \cite{2024arXiv241209452W} and \cite{2025ApJ...982...28L}, we incorporate a sink particle module into Athena++, based on the methods of \cite{federrath2010modeling} and \cite{howard2016simulating}, to model the formation and evolution of stars. At each timestep, gas cells are evaluated to determine whether they meet the criteria for sink particle formation, which approximates the collapse of molecular gas clouds. These criteria include: Jeans instability, converging gas flow, location at the center of a dense clump, and non-overlap with existing sink or star particles. Once formed, a sink particle accretes gas from nearby cells. The accreted mass is converted into stellar mass at a rate of $20\%$ per local free-fall time. \citet{2024arXiv241209452W} demonstrated that this star formation efficiency yields a surface SFR-cold gas surface density relation consistent with that of starburst galaxies exhibiting efficient star formation. 

When the accumulated stellar mass of a sink particle exceeds a mass threshold $M_{\mathrm{cnv}}$, the minimum stellar mass required for a sink particle to be converted into a star particle in our simulations, it halts further accretion and transitions into a star particle, initiating stellar feedback. We adopt $M_{\mathrm{cnv}} = 4 \times 10^4$ M$_\odot$ for galaxies at $z \sim 1$ \citep[as in][]{2025ApJ...982...28L}, and $8 \times 10^4$ M$_\odot$ for those at $z \sim 2$, reflecting the higher central densities in $z \sim 2$ disks. Each star particle has a control volume with a radius three times that of the local grid cell size. In our simulations, a ``star particle'' represents not a single star, but a stellar cluster.

We include multiple forms of stellar feedback in simulations, including radiation pressure and heating, stellar winds, and core-collapse supernovae (CCSNe). We update the local UV flux from star particles at each time step and model radiation pressure feedback as a momentum source following \citet{2024arXiv241209452W}. Photoionization is approximated via photoelectric heating, as described in the cooling and heating section. Stellar winds are implemented based on the metallicity-dependent model of \citet{Jorick2021}. For each star particle, we generate a population of virtual massive stars following the initial mass function (IMF) of \citet{kroupa2001variation}. We compute the total mass loss ($m_{\rm wind}$), metal mass loss ($m_{\rm wind,metal}$), and average terminal velocity ($v_{\rm wind}$) of stellar winds from virtual massive stars at each time step. The ejected mass, momentum, and metals are then distributed to neighboring grid cells within a radius of twice the local grid size. 

Among these mechanisms, CCSNe serve as the primary drivers of galactic outflows. The virtual massive stars in each star particle act as supernova progenitors. The mass and expected lifetimes of these massive stars are recorded at formation, and the ages of star particles are continuously updated during the simulation. When a virtual star exceeds its expected lifetime, it undergoes a supernova explosion. The energy ($\Delta E_{\rm sn}$), ejecta mass ($M_{\rm ej}$), and metal yields from each supernova are calculated following the model of \citet{sukhbold2016core}. At each timestep, the contributions from all CCSNe associated with a given star particle are summed to determine the total mass, energy, and metal feedback. Virtual stars that have exploded are then removed from the list.

Simulations with spatial resolution poorer than a few parsecs cannot resolve all phases of a supernova remnant's evolution. In such cases, injecting supernova energy purely as thermal energy often leads to rapid radiative cooling and ineffective feedback \citep[e.g.,][]{2012MNRAS.426..140D, 2017MNRAS.466...11R}. Given that the resolution in this work is approximately 10 pc, which is insufficient to resolve the Sedov-Taylor phase, we adopt the supernova feedback treatment described in Appendix A of \citet{2025ApJ...982...28L}, based on the subgrid model proposed by \citet{kim2017three}. 
For each supernova event, we compute the mean gas properties in the surrounding grid cells within a radius of $r_{snr}=3\Delta x$ to determine the appropriate supernova feedback injection method.

The mean gas density within this region is calculated and used to determine the shell formation mass, $M_{\mathrm{sf}}$, following \citet{kim2015momentum}. The feedback model is then determined by the ratio between the total gas mass in the feedback region, $M_{\mathrm{SNR}}$, and $M_{\mathrm{sf}}$, defined as $R_{\mathrm{M}} = M_{\mathrm{SNR}} / M_{\mathrm{sf}}$. The coupling of supernova energy, momentum, and mass to the surrounding gas is then carried out according to this ratio:

(i)For $R_{\mathrm{M}} > 1$, where the Sedov-Taylor phase is unresolved, the feedback is implemented by injecting momentum directly, ensuring conservation of both energy and momentum;
    
(ii)For $0.027 < R_{\mathrm{M}} < 1$, the supernova energy is partitioned such that $72\%$ is deposited as thermal energy and $28\%$ as kinetic energy;
    
(iii)For $R_{\mathrm{M}} < 0.027$, all the energy is injected in kinetic form.

\citet{Hopkins18} introduced minor revisions to the scheme of \citet{kim2017three} to ensure momentum and energy conservation in all regimes; we adopt this revised implementation in our simulations. The results of \citet{kim2017three} assume a fixed supernova energy of $10^{51}\rm\,erg$ per event. In contrast, the supernova energy injected in our simulations is computed using a Kroupa IMF \citep{kroupa2001variation} together with the variable explosion models of \citet{sukhbold2016core}. This yields an average energy of $\sim8.75 \times10^{50} \rm\,erg$, slightly lower than the canonical $10^{51}\rm\,erg$ that based on which \cite{kim2017three} derived the feedback model. We expect this modest difference should have minor effect on our results. 

Moreover, owing to the limited resolution, the Sedov–Taylor phase of most supernova remnants is not resolved in our simulations. In such cases, supernova feedback energy is injected primarily in kinetic form to mitigate artificial overcooling. In addition, in our galaxy-scale simulations with a spatial resolution of $\sim$10 pc, each star particle produces a series of core-collapse supernovae within a small volume. We therefore account for the effects of clustered supernovae \citep{gentry2017enhanced}, which increase the effective momentum injected per individual SN event. The collective momentum input from clustered supernovae can efficiently drive galactic outflows \citep{gentry2017enhanced}. Following \citet{2025ApJ...982...28L}, we adopt a momentum boost factor of $f_{\mathrm{boost}} = 10$ for supernova feedback.

In reality, only a fraction of the gas in giant molecular clouds (GMCs) forms stars; the rest is dispersed by feedback from young massive stars. In low-resolution simulations, neglecting this dispersal can lead to an overestimation of star formation efficiency. To better capture this effect, \citet{2025ApJ...982...28L} implemented a simplified model to mimic the return of dispersed gas from GMCs back to the diffuse interstellar medium (ISM). Following this approach, we return a portion of gas from each sink particle to the surrounding grid cells, based on $M_{\mathrm{gas,ini}}$, the initial gas mass of the particle at the time it transitions from a sink to a star particle:
\begin{equation} 
f_{\text{return}} =
\begin{cases}
\left( \dfrac{M_{\mathrm{gas,ini}}}{10^4\, M_\odot} \right)^\beta, & \text{if } M_{\mathrm{gas,ini}} > 10^4\, M_\odot \\
1, & \text{if } M_{\mathrm{gas,ini}} \leq 10^4\, M_\odot
\end{cases}
\end{equation}

The dispersed gas is injected into surrounding cells within a two-cell radius over a timescale of $\tau_{\mathrm{return}}$ at the velocity of the host star or sink particle, and with a temperature of 8000 K. In our fiducial simulations, gas return from GMCs to the diffuse ISM is disabled. However, to assess its impact, we enable gas return in two additional simulations, using $\beta = -0.798$ and $\tau_{\mathrm{return}}$ = 10 Myr, while keeping all other setup parameters identical to the corresponding fiducial simulations. 

\subsection{Galaxy Models of Simulations}\label{subsec:Galaxy Models}

\begin{table*}[htbp]
    \centering
    \setlength{\tabcolsep}{1pt}
    \caption{Parameters of the models for 14 isolated disc galaxies used in this work. $M_{\mathrm{vir}}$ and $R_{\mathrm{vir}}$ are the mass and radius of dark matter halo, respectively. $c$ is the halo concentration parameter. $M_{\mathrm{bulge}}$ and $R_{\mathrm{bulge}}$ are the mass and radius of stellar bulge, respectively. $M_{\mathrm{disk}}$ denotes the combined mass of $M_{\mathrm{star}}$ and $M_{\mathrm{gas}}$. $R_{\mathrm{star}}$ ($R_{\mathrm{gas}}$) and $H_{\mathrm{star}}$ ($H_{\mathrm{gas}}$) indicate the scale length and height of stellar (gas) disc. Models labeled as `Z2' adopt smaller gas and stellar disks and a more compact halo than those labeled as `Z1', corresponding to initial conditions designed to represent isolated disc galaxies at $z \sim 2$ and $z \sim 1$, respectively.} In particular, the model `Z2\_C' represents a more compact galaxy disc.
    \renewcommand{\arraystretch}{1.1}
    \begin{tabular*}{\textwidth}{@{\extracolsep{\fill}}l*{12}{c}}
    \toprule
        {Name} & \shortstack{ $M_{\mathrm{vir}}$ \\ ($10^{10}$ M$_\odot$)} & \shortstack{ $R_{\mathrm{vir}}$ \\ (kpc)} & \shortstack{ c \\ ~ } & \shortstack{ $M_{\mathrm{bulge}}$ \\ ($10^{8}$ M$_\odot$) } & \shortstack{ $R_{\mathrm{bulge}}$ \\ (kpc) } & \shortstack{ $M_{\mathrm{disk}}$ \\ ($10^{10}$ M$_\odot$) } & \shortstack{ $M_{\mathrm{star}}$ \\ ($10^{9}$ M$_\odot$) } & \shortstack{ $R_{\mathrm{star}}$ \\ (kpc) } & \shortstack{ $H_{\mathrm{star}}$ \\ (kpc) } & \shortstack{ $M_{\mathrm{gas}}$ \\ ($10^{9}$ M$_\odot$) } & \shortstack{ $R_{\mathrm{gas}}$ \\ (kpc) } & \shortstack{ $H_{\mathrm{gas}}$ \\ (kpc) } \\
        \midrule
        Z1F85D & 8.0 & 60 & 6 & 4.17 & 0.15 & 1.15 & 1.38 & 0.8 & 0.15 & 10.2 & 1.1 & 0.18\\
        Z1F70D & 8.0 & 60 & 6 & 4.17 & 0.15 & 1.15 & 3.53 & 0.8 & 0.15 & 8.0 & 1.1 & 0.18\\
        Z1F70DG & 8.0 & 60 & 6 & 4.17 & 0.15 & 1.15 & 3.53 & 0.8 & 0.15 & 8.0 & 1.1 & 0.18\\
        Z1F50D & 8.0 & 60 & 6 & 4.17 & 0.15 & 1.15 & 5.55 & 0.8 & 0.15 & 5.97 & 1.1 & 0.18\\
        Z1F50DG & 8.0 & 60 & 6 & 4.17 & 0.15 & 1.15 & 5.55 & 0.8 & 0.15 & 5.97 & 1.1 & 0.18\\
        Z1F50*S & 8.0 & 60 & 6 & 4.17 & 0.15 & 0.75 & 3.53 & 0.8 & 0.15 & 3.95 & 1.1 & 0.18\\
        Z1F30D & 8.0 & 60 & 6 & 4.17 & 0.15 & 1.15 & 7.94 & 0.8 & 0.15 & 3.58 & 1.1 & 0.18\\
        Z1F30*S & 8.0 & 60 & 6 & 4.17 & 0.15 & 0.52 & 3.53 & 0.8 & 0.15 & 1.71 & 1.1 & 0.18\\
        Z1F30*XS & 8.0 & 60 & 6 & 4.17 & 0.15 & 0.22 & 1.38 & 0.8 & 0.15 & 0.77 & 1.1 & 0.18\\
        Z2F85D & 8.0 & 50 & 4.5 & 3.5 & 0.11 & 1.14 & 1.50 & 0.5 & 0.11 & 9.9 & 0.7 & 0.13\\
        Z2F70D & 8.0 & 50 & 4.5 & 3.5 & 0.11 & 1.14 & 3.20 & 0.5 & 0.11 & 8.2 & 0.7 & 0.13\\
        Z2F50D & 8.0 & 50 & 4.5 & 3.5 & 0.11 & 1.14 & 5.55 & 0.5 & 0.11 & 5.9 & 0.7 & 0.13\\
        Z2F30D & 8.0 & 50 & 4.5 & 3.5 & 0.11 & 1.14 & 7.87 & 0.5 & 0.11 & 3.53 & 0.7 & 0.13\\
        Z2\_C & 8.0 & 50 & 4.5 & 5.0 & 0.1 & 1.45 & 4.0 & 0.4 & 0.1 & 10.5 & 0.5 & 0.12\\
        \bottomrule
    \end{tabular*}
    \label{tab:initial_conditions}
\end{table*}

We construct our disk galaxy models based on observational surveys of high-redshift star-forming galaxies, including the KMOS Redshift One Spectroscopic Survey (KROSS; \citealt{stott2016kmos, harrison2017kmos}) and the MOSFIRE Deep Evolution Field Survey (MOSDEF; \citealt{kriek2015mosfire}). These surveys provide key constraints on properties such as stellar mass, disk scaling relations, and star formation thresholds for galaxies at $z>1$. 

Each galaxy is assumed to reside in a dark matter (DM) halo and to comprise a stellar bulge, stellar disk, and gas disk. The DM halo has a mass of 8 $\times 10^{10}$ M$_\odot$ and follows a Navarro-Frenk-White (NFW) profile (\citealt{navarro1996structure}) with concentration parameters $c$ = 6 and 4.5 at $z \sim$ 1 and 2, respectively (\citealt{dutton2014cold}). The stellar bulge is modeled with a mass of $M_{\mathrm{bulge}}=4.17 \times 10^{8}$ M$_\odot$ and 3.5 $\times 10^{8}$ M$_\odot$ at $z \sim$ 1 and 2, and follows a King profile (\citealt{king1966structure}):
\begin{equation}
\rho_{\mathrm{King}}(r) = \frac{\rho_0}{1 + \frac{1}{2}\left(\frac{r}{r_{\mathrm{bulge}}}\right)^3}
\end{equation} 
where $r_{\mathrm{bulge}}$ is set to $20\%$ of the stellar disk scale length. 

The gas disk masses and gas fractions in our models are set following \citealt{2020ARA&A..58..157T}, and the resulting rotation curves are validated against observations, such as those from \citealt{2020ApJ...902...98G}. At each redshift, our disk galaxies vary in the stellar-to-gas mass ratio, characterized by the gas fraction $f_{\mathrm{gas}} = {M_{\mathrm{gas}}}/({M_{\mathrm{gas}} + M_{*}})$, where $M_{*}$ includes the stellar disk and bulge. At $z \sim$ 1, our fiducial galaxies have a total disc (gas + star) mass of $1.15 \times 10^{10}$ M$_\odot$, with initial gas fractions of 30, 50, 70 and 85 percent. These models are labeled as `FXXD', where `XX' denotes the gas fraction and `D' indicates the default total disk mass.

In addition, we include two additional models (`F50S' and `F30S') that match the stellar mass of `F70D' but adopt lower gas fractions of $50\%$ and $30\%$. Another model, `F30*XS', shares the stellar mass of `F85D' but has a gas fraction of $30\%$. Here, `*S' and `*XS' denote small and extra-small stellar masses. For galaxies at $z \sim$ 2, we focus on models with the same total disk mass but varying gas fractions. Therefore, we get a range of stellar disk masses ranging from 1.38 $\times 10^{9}$ M$_\odot$ to 7.94 $\times 10^{9}$ M$_\odot$, and initial gas disc masses from 0.77 $\times 10^{9}$ M$_\odot$ to 10.2 $\times 10^{9}$ M$_\odot$. 

Stellar disks follow the Miyamoto-Nagai profile \citep{miyamoto1975three}, while gaseous disks are modeled using a composite triple Miyamoto-Nagai profile (\citealt{smith2015simple}). The gas disk scale lengths $R_{\mathrm{gas}}$, and the scale heights $H_{\mathrm{gas}}$, are set to approximately match the redshift-dependent mass-size relation from \citealt{van20143d, ward2024evolution}. $R_{\mathrm{gas}}$ is approximately 1.4--1.8 times of the stellar counterpart, following empirical relations (\citealt{1997A&A...324..877B, 2016MNRAS.460.2143W, 2017A&A...605A..18C}). 

The density profile of the rotating, isothermal gas disk in hydrostatic equilibrium is given by (\citealt{strickland2000starburst, cooper2008three}):
\begin{align}
\begin{split}
\rho(r, z) &= \rho_{0} \times\\&
\exp \left[ - \frac{\Phi_{\mathrm{tot}}(r, z) - e^2 \Phi_{\mathrm{tot}}(r, 0) - (1 - e^2) \Phi_{\mathrm{tot}}(0, 0)}{c_{\mathrm{s}}^2} \right],
\end{split}
\end{align}
where $\rho_{0}$ is the central gas density, determined by $M_\mathrm{{gas}}$, $\Phi_{\mathrm{tot}}$ is the combined gravitational potential from all mass components. At cosmic noon, gas disks are expected to exhibit higher central densities than their local counterparts. The rotational support factor $e$ accounts for the contribution of rotation to vertical support and is defined as $e = e_{rot}(-z/z_{rot})$. We adopt $e_{rot}$ = 0.90 as the fiducial value to reflect the increased vertical thickness of high-redshift gaseous disks, and set $z_{rot}$ = 5 kpc. The effective sound speed $c_s$ incorporates both thermal and turbulent pressure support. In our simulations, we adopt $c_{s} \sim 5 \times 10^6 \,\rm{cm\,s}^{-1}$, while for the halo gas, we set $c_{s,halo}=3 \times 10^7 \,\rm{cm\,s}^{-1}$. This corresponds to a hot halo temperature of approximately 2 $\times10^6$ K. 

The details of each galaxy model are summarized in Table~\ref{tab:initial_conditions}. Each simulation is named to reflect its key characteristics. For example, in the model name `Z1F50D', `Z1' denotes initial conditions resembling isolated galaxies at redshift $z \sim$1; `F50' specifies a gas fraction of $50\%$ in the galactic disk; and `D' indicates the default total disk mass. The suffixes `*S' and `*XS' refer to small and extra-small stellar disk masses relative to the default case, while `G' denotes the inclusion of gas return processes. In the model `Z2\_C', the suffix `C' signifies that the galaxy is initialized with a more compact disk. We note that all our simulations are idealized and non-cosmological. Accordingly, differences between the `Z1' and `Z2' runs reflect variations in the adopted initial disk and halo configurations, rather than true evolutionary effects with redshift.

Table~\ref{tab:sound_velocity} provides a summary of the central gas densities $\rho_0$ and effective sound speeds $c_{s}$ used in the isothermal disk setup for each simulation. All simulations are performed in an $8,\mathrm{kpc}$ cubic domain, with a spatial resolution of $16\,\mathrm{pc}$ in the central starburst region (defined as $r<1\,\mathrm{kpc}$ in the disk and $|z|<120\,\mathrm{pc}$). Outside this region, the refinement strategy progressively coarsens the resolution to 32, 64, and 128 pc to improve computational efficiency.

\begin{table}[htbp]
    \centering
    \setlength{\tabcolsep}{15pt} %
    \caption{The central densities $\rho_0$ and effective sound speed $c_s$ of gas disc. $c_s$ is determined by the combined energy density, which includes both the turbulence-induced dynamic pressure and the internal energy-derived static pressure.}
    \label{tab:sound_velocity}
    \begin{tabular}{lcc}
        \toprule
        Name & \(\rho_0\) [g cm$^{-3}$] & \(c_s\) [cm s$^{-1}$]  \\
        \midrule
        Z1F85D & 1370 & \(5.9 \times 10^6\) \\
        Z1F70D & 1100 & \(5.6 \times 10^6\) \\
        Z1F70DG & 1100 & \(5.6 \times 10^6\) \\
        Z1F50D & 830  & \(5.4 \times 10^6\) \\
        Z1F50DG & 830  & \(5.4 \times 10^6\) \\
        Z1F50*S & 570  & \(4.5 \times 10^6\) \\
        Z1F30D & 520  & \(5.3 \times 10^6\) \\
        Z1F30*S & 245  & \(3.8 \times 10^6\) \\
        Z1F30*XS & 110  & \(3.0 \times 10^6\) \\
        Z2F85D & 4330 & \(6.75 \times 10^6\) \\
        Z2F70D & 3620 & \(6.36 \times 10^6\) \\
        Z2F50D & 2600 & \(6.28 \times 10^6\) \\
        Z2F30D & 1530 & \(6.0 \times 10^6\) \\
        Z2\_C & 7300 & \(8.0 \times 10^6\) \\
        \bottomrule
    \end{tabular}
\end{table}

\begin{figure*}[htb]
\begin{centering}
\hspace{-0.0cm}
\includegraphics[width=1.0\textwidth, clip]{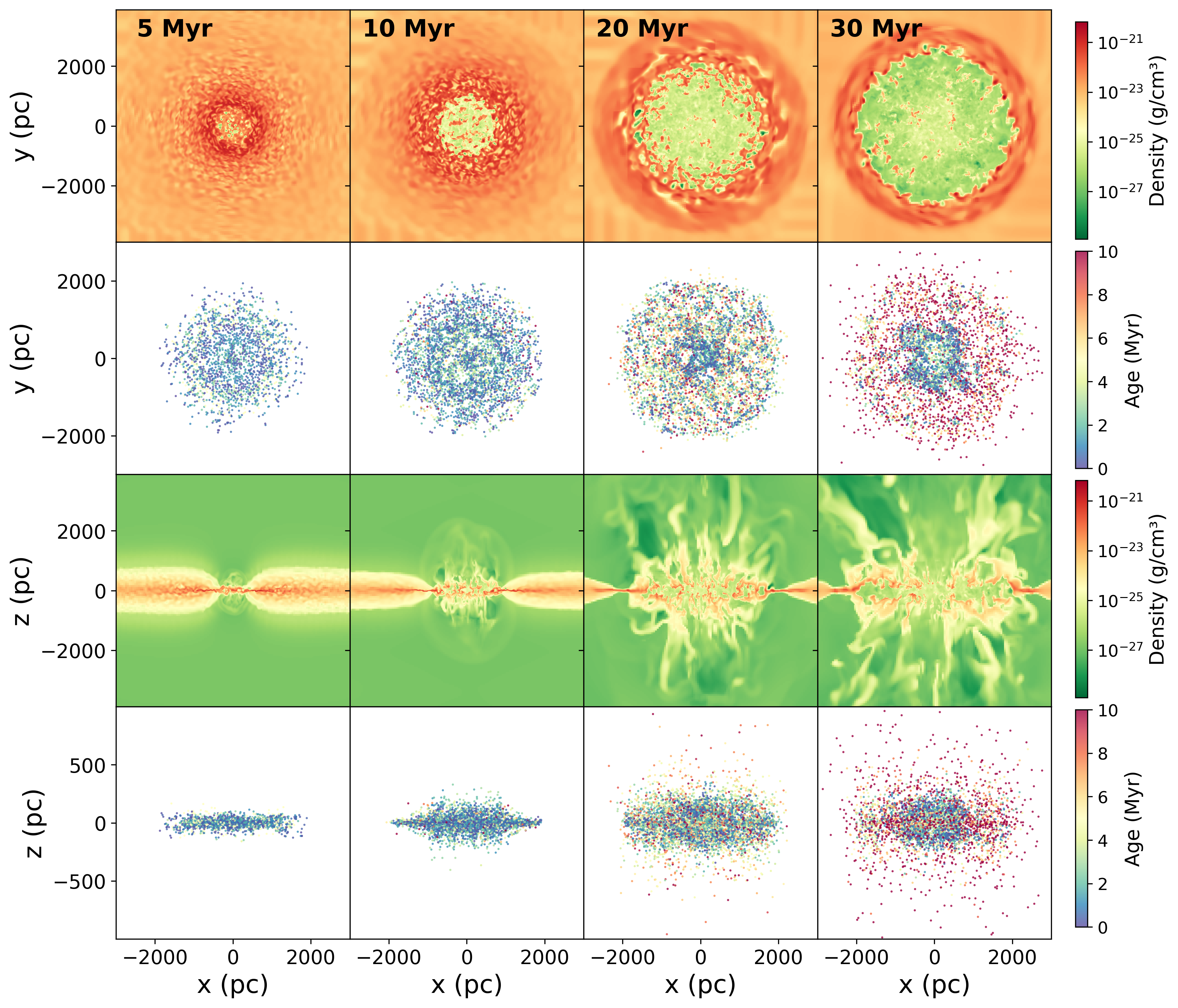}
\caption{Evolution of the gas disk and stellar particle distribution in face and edge-on views for Z1F70D at t = 5, 10, 20, and 30 Myr. The first and third rows display the evolution of the gas component, while the second and fourth rows show the spatial distribution of stellar particles over a 10 Myr interval.}
\label{fig:begain}
\end{centering}
\end{figure*}

\section{Results} \label{sec:results}
In general, our simulations produce starbursts lasting approximately 20--30 Myr, which subsequently drive the development of multiphase galactic winds. As an illustrative example, Figure~\ref{fig:begain} shows the evolution of the gas disk and the stellar particle distribution in the Z1F70D simulation. By $t=5$ Myr, a significant number of stars have formed, generating a small bubble in the central region. As star formation continues, the outflow rapidly expands. By $t=20$ Myr, the outflow has reached the boundary of the simulation domain, and by $t=30$ Myr, it occupies nearly the entire volume above the disk.

The following subsections present a detailed analysis of our results. Section~\ref{subsec:starburst} examines the characteristics of starbursts across different simulations. Section~\ref{subsec:overall view} provides an overview of galactic outflows under varying initial conditions. The quantitative properties of these outflows, including outflow velocity $V_{\mathrm{out}}$, mass outflow rate $\dot M_{\mathrm{out}}$, and mass loading factor $\eta_{\mathrm{M}}$, are discussed in Sections~\ref{subsec:outflow velocity}, \ref{subsec:mass outflow rate}, and \ref{subsec:mass loading factor}, with comparisons to observations and other simulations. Finally, Section~\ref{subsec:time evolution and radial} explores the temporal evolution and radial profiles of the outflows in our simulations. We note that our simulations adopt somewhat idealized initial conditions, which may partially deviate from the physical conditions of real galaxies. In addition, the simulation duration of 30 Myr is relatively short compared to the full evolutionary timescale of galactic outflows. As a result, comparisons between our results and observed wind properties, such as mass outflow rates and mass loading factors, may be subject to systematic uncertainties.

\begin{figure}[htb]
\begin{centering}
\hspace{-0.0cm}
\includegraphics[width=1.0\columnwidth, clip]{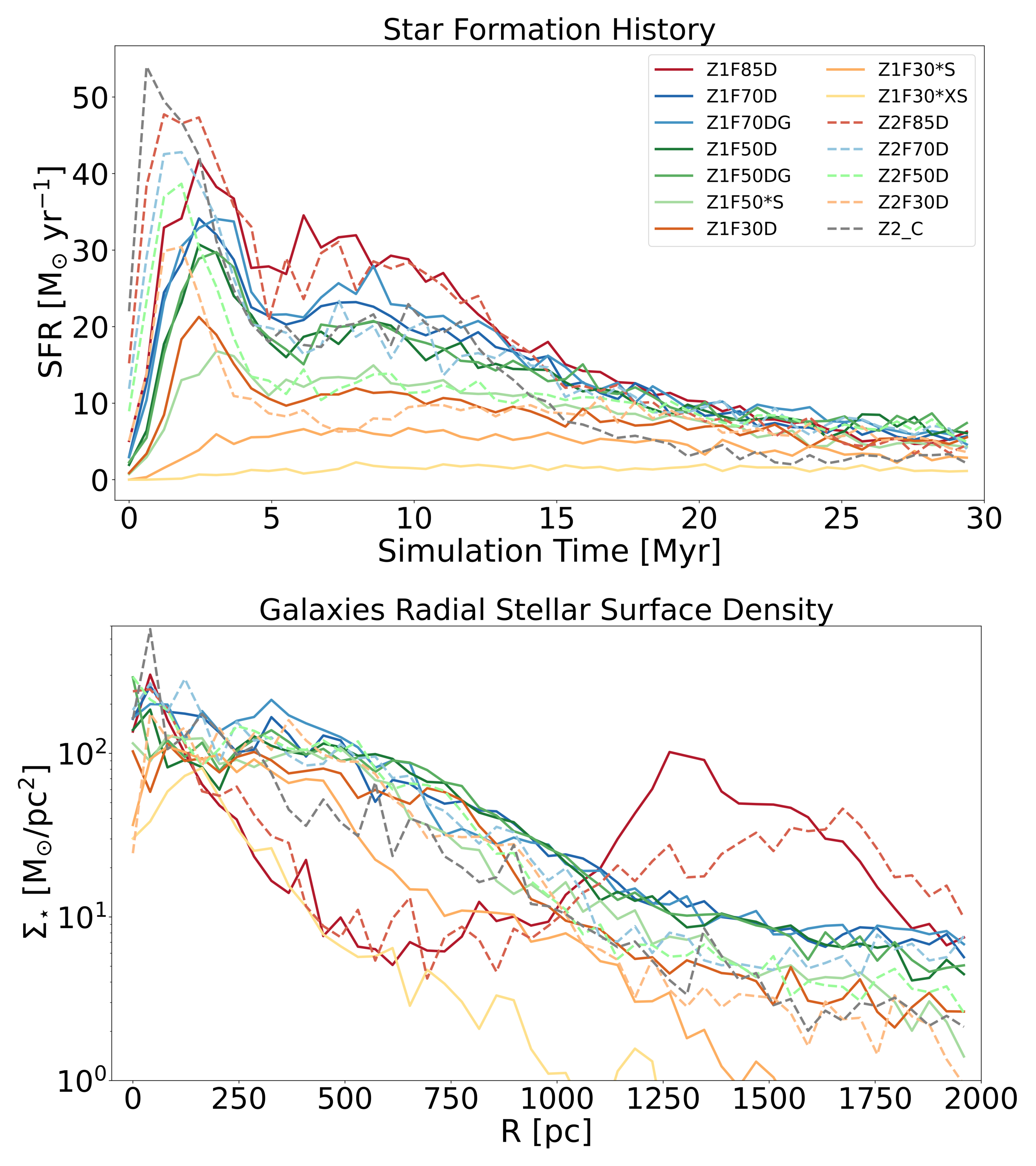}
\caption{Top: The time evolution of the star formation rate (SFR) for each simulation is shown. Solid lines denote simulations labeled `Z1', corresponding to models with initial conditions representative of galaxies at $z\sim1$, while dashed lines indicate the `Z2' simulations. Bottom: Radial surface density profiles of stars formed at 30 Myr. Simulations with an initial gas fraction of $85\%$ show a modest secondary peak at approximately 1.5 kpc.}
\label{fig:SFH_RSD}
\end{centering}
\end{figure}

\begin{figure*}[htb]
\begin{centering}
\hspace{-0.0cm}
\includegraphics[width=1.0\textwidth, clip]{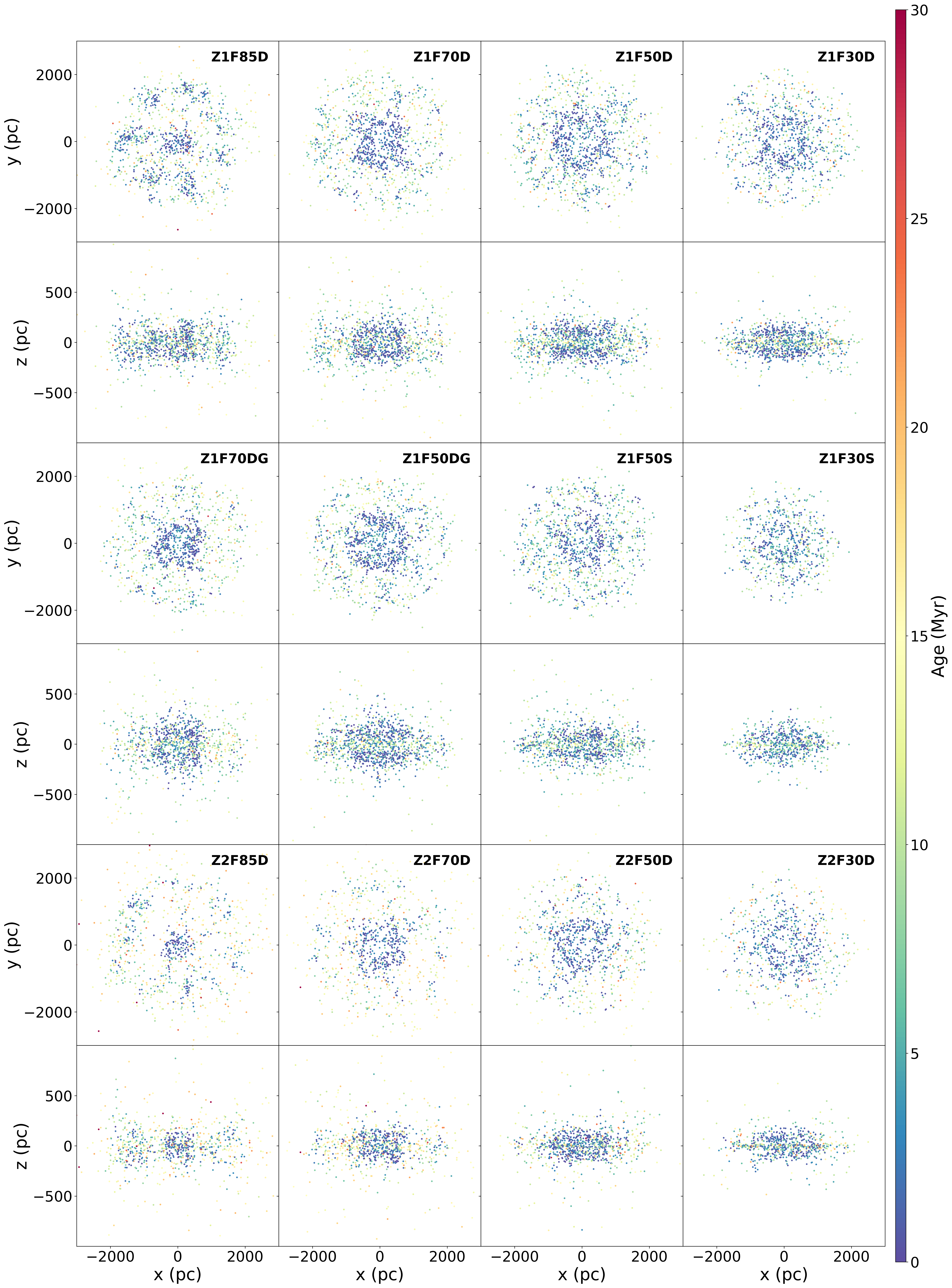}
\caption{Face-on and edge-on views of the distribution of star particles formed during the starburst. The color indicates the age of particle. Note that the mass of each stellar particle is 4 and $8 \times 10^4$ M$_\odot$ in the `Z1' and `Z2' simulations.}
\label{fig:star}
\end{centering}
\end{figure*}

\begin{table*}[htbp]
    \centering
    \caption{The SFRs and wind properties in our simulations. SFRs include the maximum instantaneous values, the average values over the 0--30 Myr, as well as the instantaneous values at 30 Myr. The outflow properties including outflow velocity, mass outflow rate, and mass loading factor for the cool, warm, and hot phases are reported at $t=30$ Myr.}
    \setlength{\tabcolsep}{1pt}
    \renewcommand{\arraystretch}{1.1} 
    \label{tab:parameters}
    \begin{tabular*}{\textwidth}{@{\extracolsep{\fill}}l*{12}{c}}
        \toprule
        {Name} 
      & \shortstack{$\mathrm{SFR}_{\mathrm{max}}$ \\ (M$_\odot \,\rm{yr}^{-1}$)}
      & \shortstack{$\overline{\mathrm{SFR}}$ \\ (M$_\odot \,\rm{yr}^{-1}$)}
      & \shortstack{$\mathrm{SFR}_{30}$ \\ (M$_\odot \,\rm{yr}^{-1}$)}
      & \shortstack{$V_{\mathrm{out,c}}$ \\ ($\,\rm{km\,s}^{-1}$)}
      & \shortstack{$V_{\mathrm{out,w}}$ \\ ($\,\rm{km\,s}^{-1}$)}
      & \shortstack{$V_{\mathrm{out,h}}$ \\ ($\,\rm{km\,s}^{-1}$)}
      & \shortstack{$\dot M_{\mathrm{out,c}}$ \\ (M$_\odot \,\rm{yr}^{-1}$)}
      & \shortstack{$\dot M_{\mathrm{out,w}}$ \\(M$_\odot \,\rm{yr}^{-1}$)}
      & \shortstack{$\dot M_{\mathrm{out,h}}$ \\ (M$_\odot \,\rm{yr}^{-1}$)}
      & \shortstack{$\eta_{\mathrm{M,c}}$ \\ ~}
      & \shortstack{$\eta_{\mathrm{M,w}}$ \\ ~}
      & \shortstack{$\eta_{\mathrm{M,h}}$ \\ ~}  \\
        \midrule
        Z1F85D & 41.07 & 17.88 & 6.53 & 113.75 & 194.13 & 618.88 & 14.71 & 2.80 & 1.67 & 2.70 & 0.51 & 0.31\\
        Z1F70D & 40.8 & 14.64 & 6.40 & 120.84 & 191.18 & 520.15 & 4.20 & 1.93 & 0.83 & 0.73 & 0.33 & 0.14\\
        Z1F70DG & 36.13 & 15.73 & 4.67 & 127.38 & 204.70 & 144.60 & 3.72 & 1.30 & 0.19 & 0.65 & 0.23 & 0.03\\
        Z1F50D & 31.87 & 13.23 & 5.87 & 169.01 & 215.07 & 453.01 & 1.05 & 1.17 & 0.61 & 0.16 & 0.18 & 0.09\\
        Z1F50DG & 30.53 & 13.50 & 6.53 & 202.10 & 319.31 & 152.26 & 1.64 & 1.18 & 0.28 & 0.22 & 0.16 & 0.04\\
        Z1F50*S & 18.0  & 9.05 & 3.73 & 197.80 & 242.68 & 505.13 & 0.78 & 0.71 & 0.60 & 0.16 & 0.15 & 0.12\\
        Z1F30D & 21.73  & 8.55 & 5.07 & 179.11 & 249.15 & 460.34 & 0.32 & 0.52 & 0.64 & 0.07 & 0.11 & 0.13\\
        Z1F30*S & 7.60  & 4.52 & 3.20 & 208.82 & 282.43 & 437.34 & 0.15 & 0.18 & 0.44 & 0.05 & 0.06 & 0.14\\
        Z1F30*XS & 2.53  & 1.32 & 1.60 & 186.97 & 203.16 & 165.74 & 0.0 & 0.02 & 0.25 & 0.0 & 0.01 & 0.16\\
        Z2F85D & 54.93 & 18.21 & 3.73 & 65.43 & 168.64 & 551.08 & 16.48 & 1.49 & 1.24 & 4.02 & 0.36 & 0.30\\
        Z2F70D & 48.27 & 15.19 & 5.07 & 62.86 & 171.44 & 576.17 & 12.93 & 1.43 & 0.91 & 2.17 & 0.24 & 0.15\\
        Z2F50D & 42.67 & 12.06 & 5.07 & 54.22 & 154.27 & 613.68 & 8.11 & 1.06 & 0.65 & 1.37 & 0.18 & 0.11\\
        Z2F30D & 34.13 & 9.16 & 4.53 & 82.36 & 178.52 & 597.72 & 1.62 & 0.92 & 0.55 & 0.29 & 0.17 & 0.10\\
        Z2\_C & 68.0 & 13.96 & 1.33 & 45.56 & 128.46 & 385.42 & 3.96 & 0.53 & 0.35 & 1.46 & 0.19 & 0.13\\
        \bottomrule
    \end{tabular*}
\end{table*}

\begin{figure*}[htb]
\begin{centering}
\hspace{-0.0cm}
\includegraphics[width=1.0\textwidth, clip]{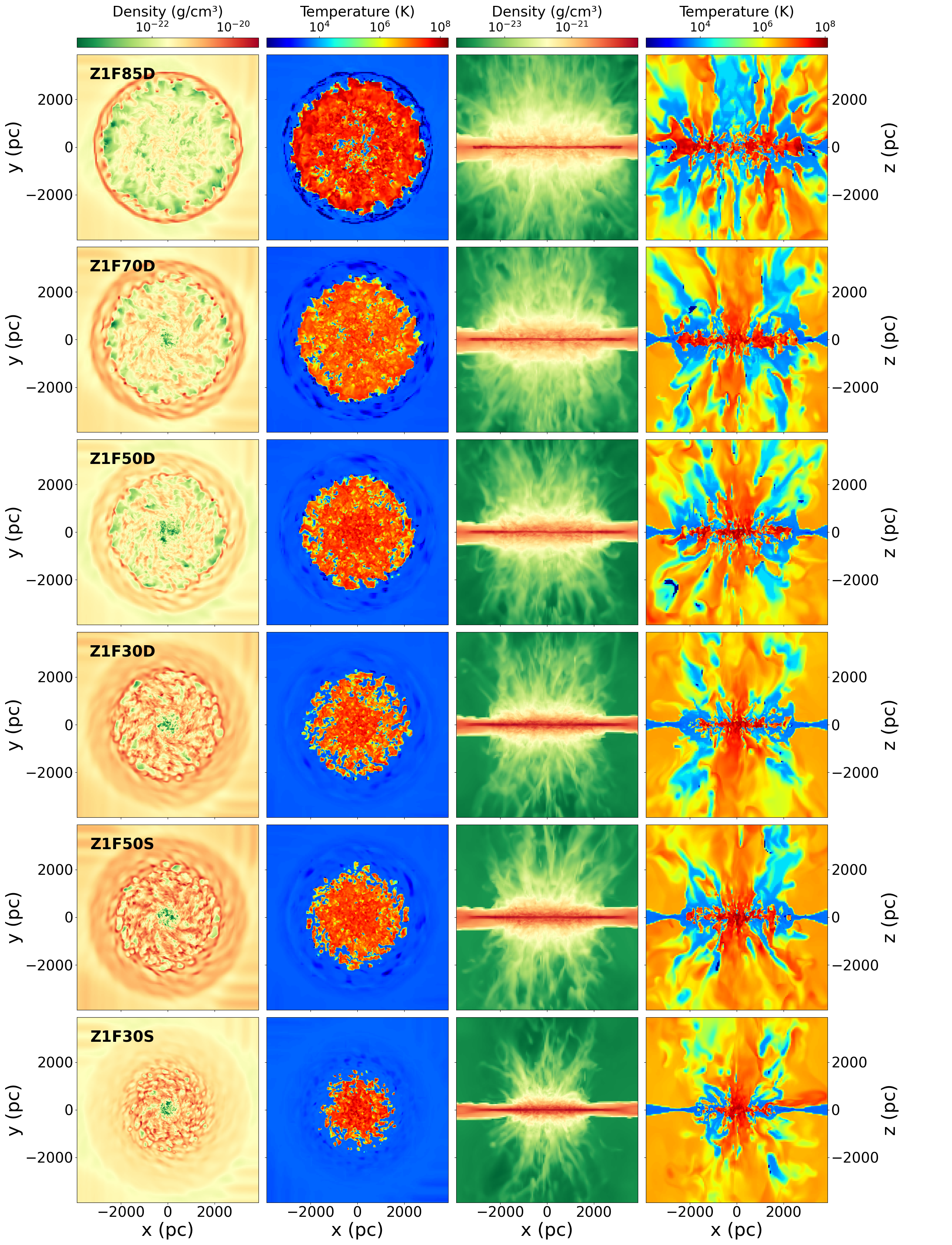}
\caption{Face-on and edge-on galaxies outflows of the `Z1' models at t = 30 Myr. From left to right, the panels show: (1) the face-on projected gas density, (2) the face-on temperature distribution, (3) the edge-on projected gas density, and (4) the edge-on temperature distribution. Simulations with higher initial gas fraction generate more spatially extended outflows with wider opening angles and more pronounced multi-phase structures.}
\label{fig:z1_530}
\end{centering}
\end{figure*}

\begin{figure}[htb]
\begin{centering}
\hspace{-0.0cm}
\includegraphics[width=1.0\columnwidth, clip]{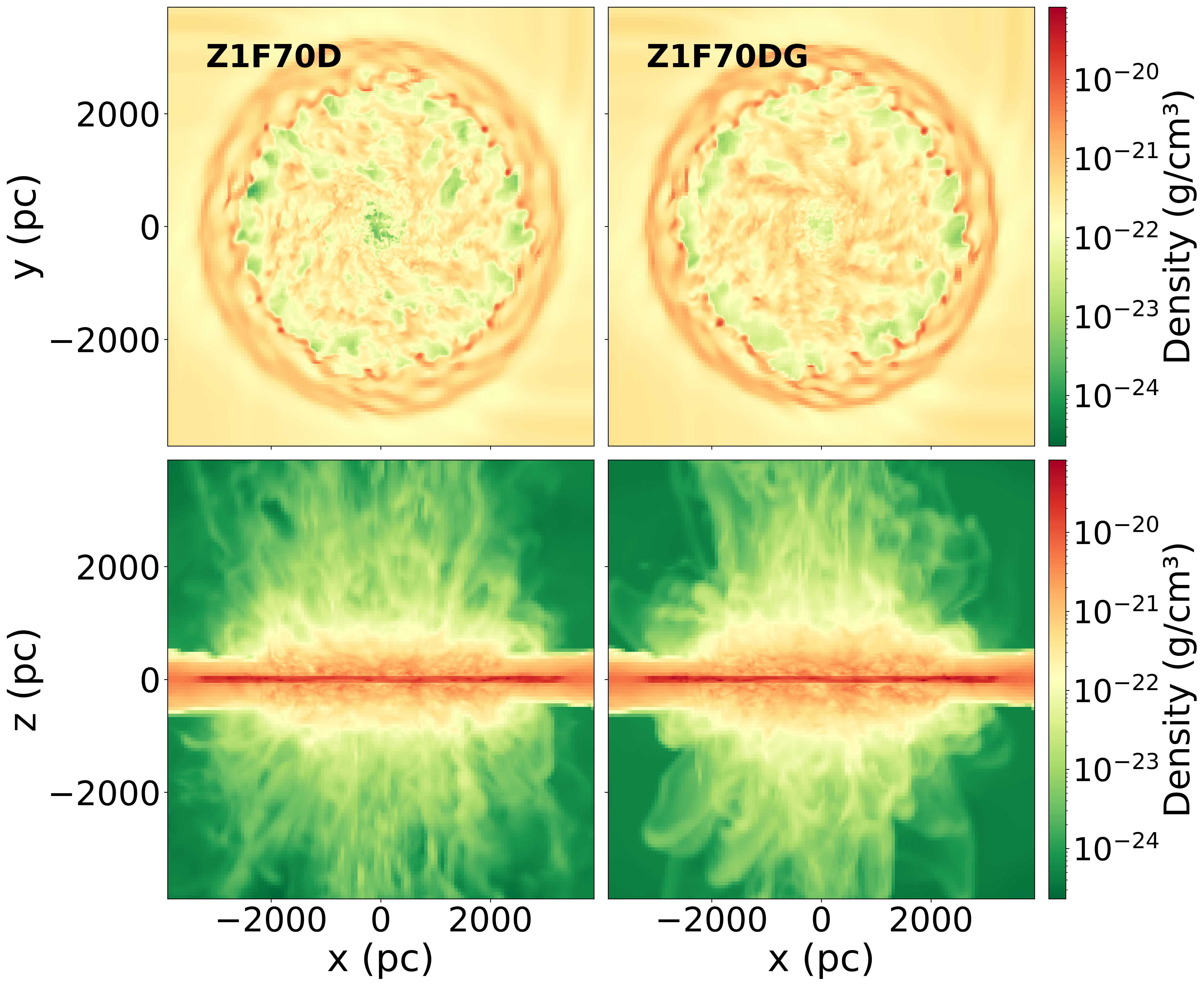}
\caption{Same as Figure~\ref{fig:z1_530} but for a direct comparison of projected gas density between Z1F70D and Z1F70DG. In order to compare the effects of the gas return process on galactic outflows under the same initial conditions.}
\label{fig:z1_gasreturn}
\end{centering}
\end{figure}

\subsection{Starburst}\label{subsec:starburst}

Figure~\ref{fig:SFH_RSD} shows the time evolution of the star formation rate (SFR) in our simulations. The SFR rises to peak values of 2--68 M$_\odot \,\rm{yr}^{-1}$ within the first 1--8 Myr, then declines due to stellar feedback and the depletion of gas available for collapse, reaching 2--6 M$_\odot \,\rm{yr}^{-1}$ by the end of the simulations. The peak star formation rate and its timing depend on the initial gas fraction, total gas disk mass, and disk compactness. Higher initial gas fractions produce larger SFR peaks, while more compact disks drive more intense starbursts due to higher gas column densities and more rapid gravitational collapse. In some galaxies, the star formation history (SFH) shows sharp fluctuations with time scale of $\sim$1--2 Myr.

Simulations with initial gas fractions above $50\%$ exhibit two distinct phases of elevated star formation. In the first $\sim$5 Myr, the SFR rises sharply due to rapid gravitational collapse in the central dense regions, aided by efficient radiative cooling. This is followed by a quick decline. Between 5 and 15 Myr, the SFR increases again moderately and stays above 10 M$_\odot \,\rm{yr}^{-1}$ for roughly 10 Myr. The second phase likely results from the cooling and collapse of large-scale dense clumps in the outer disk, combined with cumulative feedback effects from early supernovae and radiation pressure that compress the surrounding ISM. In contrast, disks with lower initial gas fractions lack sufficient cool gas to sustain a second starburst. Their SFHs typically show a single peak followed by a plateau, with SFRs remaining at modest to low levels through the remainder of the simulation. 

Simulations initialized to resemble galaxies at 
$z\sim2$ (labeled `Z2') exhibit higher peak SFRs than the `Z1' cases, primarily because their disks are set to be more compact and have higher gas column densities, as motivated by observational constraints and empirical scaling relations. However, the more rapid consumption of star-forming gas in these systems leads to a faster post-peak decline in the SFR. We again emphasize that our simulations model idealized, isolated galaxies with with initial conditions designed to resemble systems at $z\sim1$ and $z\sim2$, but do not capture the full cosmological evolution of galaxies. Additionally, the star formation histories of Z1F70DG and Z1F50DG closely resemble those of Z1F70D and Z1F50D, respectively, suggesting that gas return has only a modest impact on the overall starburst behavior.

To explore whether more intense starbursts can be triggered within the same physical framework, we constructed a more compact initial gas disk in the `Z2\_C' simulation. As shown in Figure~\ref{fig:SFH_RSD}, Z2\_C exhibits a rapid rise in SFR, peaks at 68.0 M$_\odot \,\rm{yr}^{-1}$ around $t \sim 1$ Myr---the highest among all simulations---followed by a steep decline to 3.2 M$_\odot \,\rm{yr}^{-1}$ at 20 Myr, and a gradual decrease to approximately 1 M$_\odot \,\rm{yr}^{-1}$ by 30 Myr. This result demonstrates that our framework can reproduce more intense starbursts when the gas disk is more compact. Table~\ref{tab:parameters} summarizes the peak, average, and final SFRs to quantify these trends.

In addition to differences in SFR, the spatial distribution of newly formed stars also varies across simulations, potentially affecting outflow properties. As shown by the bottom panel of Figure~\ref{fig:SFH_RSD}, higher gas fraction models exhibit stronger central stellar surface densities, consistent with the compact stellar clumps seen in Figure~\ref{fig:star} within $r < 2$ kpc and $|z| < 250$ pc---regions of highest gas density. These stars form in clumps without clear spiral structures, consistent with observations \citep{2011ApJ...733..101G, 2014ARA&A..52..291C, 2015ApJ...800...39G, 2020ARA&A..58..661F}.

A higher initial gas fraction leads to more widespread and enhanced star formation across the disk. For example, Z1F70D, Z1F50*S, and Z1F30*S share the same stellar disk mass, but decreasing gas fractions progressively limit the radial extent of star formation. In simulations with $f_{\mathrm{gas}} = 85\%$, Z1F85D and Z2F85D, clustered stellar structures also form in the outer disk, as seen by the secondary bump at $\sim$1.5 kpc in the bottom panel of Figure~\ref{fig:SFH_RSD} and in Figure~\ref{fig:star}.

In simulations with the gas return process (Z1F70DG and Z1F50DG), young stars in central regions are more vertically extended, although their surface densities remain similar to those without gas return. This likely reflects enhanced cold and cool gas outflows facilitating star formation at higher altitudes.

The bottom panel of Figure~\ref{fig:SFH_RSD} further shows that the spatial distribution of newly formed stars in simulations designed to represent galaxies at $z\sim2$ is broadly similar to that at 
$z\sim1$, with higher gas fractions producing more extended stellar structures. Note that, the larger mass per star particle in `Z2' runs reduces the the total number of particles, which can visually amplify differences relative to `Z1' runs in Figure~\ref{fig:star}.

\subsection{Overall View of Winds} 
\label{subsec:overall view}

Multi-wavelength observations reveal that galactic winds are intrinsically multiphase, comprising hot X-ray, emitting plasma, warm and cool gas traced by optical/UV lines, and cold gas seen in the radio (e.g., \citealt{2017arXiv170109062H, 2020A&ARv..28....2V}). In all our simulations, starbursts generate galactic-scale multiphase outflows, but their morphology and properties vary with the initial disc conditions. Figure~\ref{fig:z1_530} shows face-on and edge-on views of projected gas density and temperature slices at $t=30$ Myr for six simulations modeling galaxies at $z\sim1$. Face-on views reveal a hot, low-density central bubble formed by sustained supernova feedback, surrounded by cool gas clumps and a dense ring likely formed by bubble-driven compression. Edge-on views highlight the characteristic biconical, multiphase structure of the outflows.

At fixed total disk mass, higher gas fractions lead to outflows with wider opening angles and larger volumes. The central hot bubble also expands with increasing initial gas fraction, driven by more vigorous star formation and supernova feedback. When the initial gas mass is similar but stellar masses differ, such as in Z1F30D and Z1F50*S, the resulting outflow morphology and stellar distributions remain comparable. 

Figure~\ref{fig:z1_gasreturn} further compares the projected gas density in Z1F70D and Z1F70DG, showing that the gas return process leads to more collimated outflows with narrower biconical angles, consistent with the findings of \cite{2025ApJ...982...28L}. Meanwhile, simulations modeling galaxies at $z\sim2$ exhibit wind morphologies broadly similar to those at $z\sim1$, but with somewhat more extended volumes, which should mainly result from the higher initial gas densities and more compact disc. A visual comparison between the `Z1' and `Z2' runs is presented in Appendix~\ref{app:redshift}.

\subsection{Outflow Velocity}\label{subsec:outflow velocity}

Outflow velocity, mass outflow rate, and mass loading factor, along with their dependence on stellar mass and SFR, are key diagnostics of galactic winds, reflecting the dynamical influence of stellar feedback \citep{murray2005maximum, 2005ARA&A..43..769V, 2017arXiv170109062H}. We investigate these properties in this and subsequent subsections. In our analysis, outflowing gas is defined as gas in cells with positive radial velocity ($V_{\mathrm{out}} > 0$), excluding material within $\pm\,200$ pc of the disk midplane. Although we adopt a minimal velocity threshold of $V_{\mathrm{out}} > 0$, the vast majority of outflowing gas in our simulations has velocities exceeding 100 km/s. The outflowing gas is classified into four thermal phases: hot ($T > 5 \times 10^5$ K), warm ($2 \times 10^4 < T \le 5 \times 10^5$ K), cool ($8 \times 10^3 < T \le 2 \times 10^4$ K), and cold ($T \le 8 \times 10^3$ K). Throughout this work, the subscripts `c', `w', and `h' denote the cool, warm, and hot gas phases, respectively, as defined by our temperature criteria. We divide the outflow volume into radial shells and compute the volume-averaged $V_{\mathrm{out}}$ of each phase in each shell.

\begin{figure*}[htb]
\begin{centering}
\hspace{-0.0cm}
\includegraphics[width=0.95\textwidth, clip]{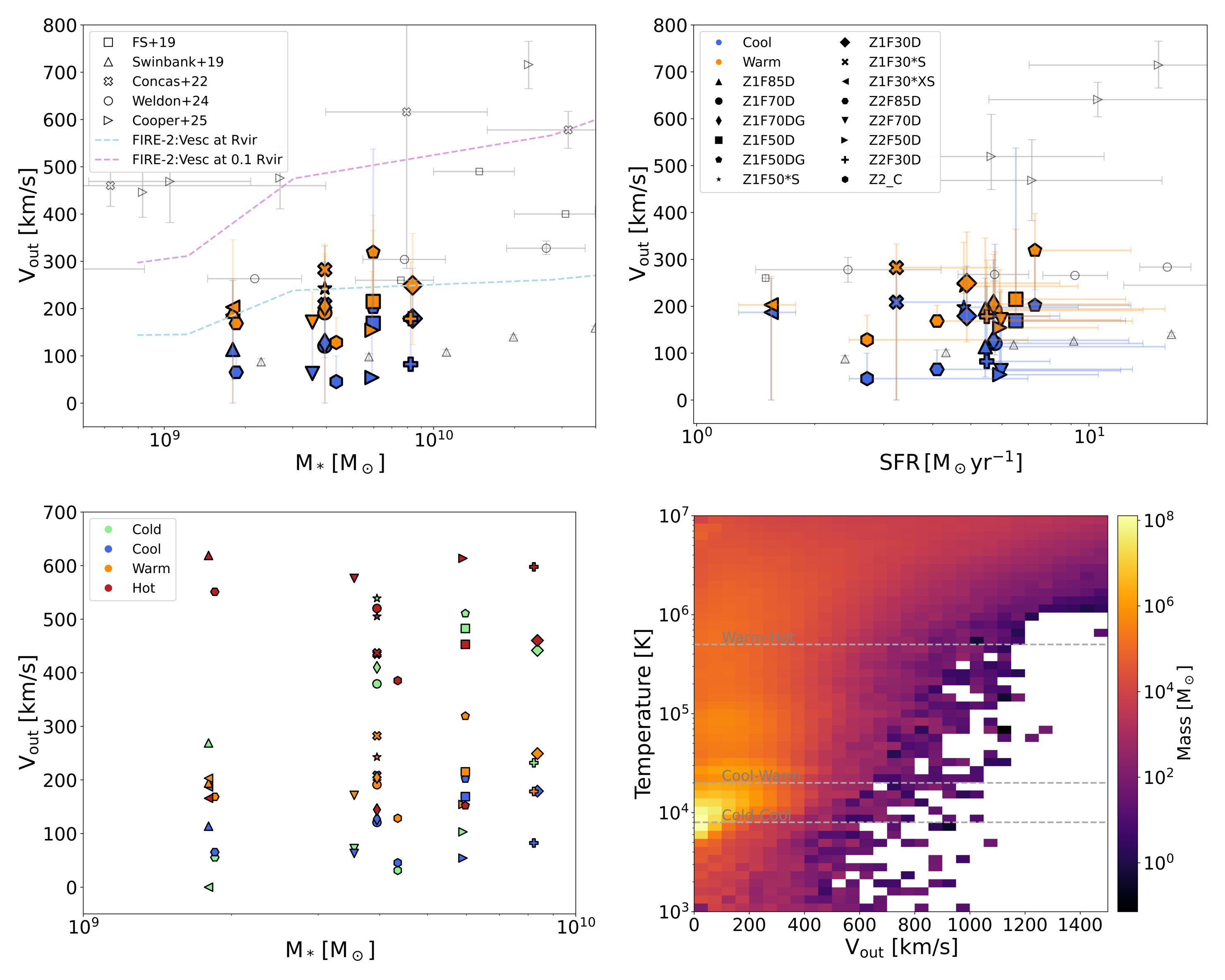}
\caption{Top: Outflow velocity as a function of stellar mass and SFR for the cool (green) and warm (orange) phases.
Middle: Same as the top panels, but for the cold (blue) and hot (red) phases.
Bottom: Distribution of outflowing gas in velocity–temperature space at $0.05R_{\mathrm{vir}}$ and $t=30$ Myr for the Z1F70D simulation. Colors indicate the gas mass in each velocity–temperature bin, and the inset shows the corresponding one-dimensional velocity PDF. Marker size indicates gas fraction. Hatched symbols denote `Z2' models, while solid symbols denote `Z1' models. The corresponding error bars indicate the maximum and minium values during the 18--30 Myr interval. The gray open squares are at 0.6 $< z <$ 2.7 from KOMS(\citealt{schreiber2019kmos3d}). The gray open inverted triangles represent starburst-driven outflows at $z \sim$ 1 from KMOS (\citealt{swinbank2019energetics}). The gray open diamonds are low-mass galaxies from \citealt{concas2022being} at cosmic noon. Gray open pentagons indicate warm ionised outflows at $z =$ 1.4--3.8 reported by \citealt{2024MNRAS.531.4560W}. Open right triangles stand for high velocity outflows at 2.5 $< z <9$ in \citealt{cooper2025high}.  The plum and light blue dashed lines show the escape velocity of gas at 0.1$R_{\mathrm{vir}}$ and $R_{\mathrm{vir}}$ obtained in the FIRE-2 simulations \citealt{pandya2021characterizing}.}
\label{fig:com}
\end{centering}
\end{figure*}

We evaluate outflow properties, include velocity, on a spherical shell of width $0.1$ kpc at a radius of 0.05$R_{\mathrm{vir}}$ (corresponding to $\sim$3 kpc and 2.5 kpc for the `Z1’ and `Z2’ models, respectively). At this radius, the multiphase outflow, including hot wind, has largely decoupled from the disc and varies relatively smoothly with radial distance. This choice is broadly consistent with previous simulations and observations. For example, \citet{porter2024any} adopted 0.05$R_{\mathrm{vir}}$ for high-redshift galaxies to minimize contamination from turbulent disc gas. Observationally, broad emission-line outflows (e.g., H$\alpha$ or $\mathrm{O,III}$) are typically detected out to radii of $\sim$1–5 kpc \citep[e.g.,][]{2005ApJS..160..115R, 2012ApJ...760..127M, newman2012sins, schreiber2019kmos3d} or to the effective radius $R_e$ \citep[e.g.,][]{1985Natur.317...44C, 2015ApJ...809..147H, llerena2023ionized, 2025MNRAS.tmp.1579S}. 

We note, however, that observations integrate emission along the line of sight, whereas simulations measure mass and energy fluxes through a shell at a fixed radius. This methodological difference may introduce systematic uncertainties when comparing simulations with observations. In addition, the outflow properties in our simulations are measured beyond $\pm$200 pc from the disk midplane. In contrast, observational studies at $z \sim$ 1--3 often lack the spatial resolution to resolve the vertical structure of disc accurately \citep{swinbank2019energetics, schreiber2019kmos3d, concas2022being, llerena2023ionized, 2024MNRAS.531.4560W, cooper2025high}. Therefore, results in observations would contain gas in the disc. This difference may contribute partly to the discrepancies between our simulations and observations. It is also important to note that the geometry of the sampled outflow region, whether spherical or conical, can significantly influence the derived mass and velocity profiles. Appendix~\ref{app:multiangle} briefly explores the impact of varying opening angles on our measurements.

The overall outflow properties at the end of our simulations are summarized in Table~\ref{tab:parameters}. Figure~\ref{fig:com} compares volume-weighted mean outflow velocities measured at $r = 0.05\,R_{\rm vir}$ and $t = 30$ Myr in our simulations with observational data. Simulation error bars indicate the velocity range over 18–30 Myr, during which the outflows evolve rapidly. Gas phases, cold, cool, warm, and hot, are shown in blue, green, orange, and red, respectively, while observations are plotted in gray with distinct symbols. The top and middle panels present outflow velocity as a function of stellar mass (left) and SFR (right) for the cool/warm and cold/hot phases, respectively. The bottom panel shows the two-dimensional distribution of outflow velocity versus temperature for model Z1F70D at $0.05\,R_{\rm vir}$ and 30 Myr, color-coded by gas mass. 

To better match observational estimates, often based on H$\alpha$ emission lines, we compute the SFR as an average over the preceding 5 Myr at each time $t$. Since H$\alpha$ traces ionizing photons from massive stars with lifespans of $\lesssim$10 Myr, this timescale reflects recent star formation activity \citep{1998ARA&A..36..189K, 2011ApJ...741..124H, 2011ApJ...737...67M, calzetti2012starformationrateindicators, 2021MNRAS.501.4812F}. 

At $t = 30$ Myr, the volume-weighted mean outflow velocities measured at $r = 0.05R_{\mathrm{vir}}$ across our simulations span $5-550\,\rm{km\,s}^{-1}$ (cold), 40--200 $\,\rm{km\,s}^{-1}$ (cool), 130--290 $\rm{km\,s}^{-1}$ (warm), and 170--1000 $\,\rm{km\,s}^{-1}$ (hot). These values are broadly consistent with observations at cosmic noon (e.g., \citealt{schreiber2019kmos3d}, \citealt{swinbank2019energetics}, \citealt{llerena2023ionized}, \citealt{2024MNRAS.531.4560W}) and at higher redshift \citep{2023arXiv231006614X}, though they are up to $\sim 50\%$ lower than in some studies (e.g., \citealt{concas2022being}, \citealt{cooper2025high}), which may be biased toward higher velocities due to limitations of medium-resolution spectra. Notably, the warm-phase outflow velocity in our simulations is comparable to the escape velocity at $R_{\mathrm{vir}}$ in FIRE-2 simulations.

Outflow velocities in our simulations generally increase with gas temperature. However, in roughly half of the runs, the volume-weighted mean velocity of the cold phase at $0.05R_{\mathrm{vir}}$ and $t=30$ Myr exceeds that of the cool phase, reaching $400$–$550\,\mathrm{km\,s^{-1}}$. As shown by the PDF in Figure~\ref{fig:com}, a subset of the cold gas attains velocities of $400$–$800\,\mathrm{km\,s^{-1}}$, likely through acceleration by interactions and mixing with the warm and hot phases \citep{2022ApJ...924...82F, schneider2024cgols, 2024arXiv241209452W}. Meanwhile, a substantial fraction of the cool phase remains at velocities below $200\,\mathrm{km\,s^{-1}}$. A likely explanation is that the cool phase dominates the outflow in mass, so multiphase interactions and mixing are insufficient to accelerate all cool gas to high velocities. In contrast, these mixing can more efficiently accelerate a substantial fraction of the colder, denser gas. Consequently, in some simulations the mean outflow velocity of the cool phase is lower than that of the cold phase at $t=30$ Myr. Moreover, different gas phases in the outflow evolve at different paces, causing their velocity peaks at a given radius to occur at different epochs (\citealt{2025ApJ...982...28L}). 

Considering the full scatter, our simulations show no clear correlation between outflow velocity and either $M_*$ or SFR. Observational results are similarly very mixed: some studies find no correlation (e.g., \citealt{schreiber2019kmos3d}, \citealt{2024MNRAS.531.4560W}, \citealt{cooper2025high}), while others report weak positive correlations with $M_*$ and/or SFR (e.g., \citealt{swinbank2019energetics}, \citealt{llerena2023ionized}).

Given the substantial scatter, correlations between outflow velocity and galaxy properties should be interpreted with caution. Several factors likely contribute to the discrepancies among observational studies. First, sample sizes are often limited. This issue is exacerbated by the fact that observations typically capture outflows at a single snapshot in time, whereas outflows evolve over tens of Myr, during which their properties can vary significantly.
Second, velocity estimates carry large uncertainties due to methodological limitations. Most observations at cosmic noon probe the warm ionized phase ($\sim10^4-10^5$ K), using emission/absorption lines such as H$\alpha$, $\mathrm{C\,II}$, $\mathrm{Fe\,II}$,$\mathrm{O\,II}$ $\mathrm{O\,III}$. A common approach is to assume that outflows contribute to the broad spectral component and to apply double Gaussian fitting to separate it from the narrow component.

However, this method faces two key challenges. First, the broad component may include contributions from disk inclination, beam smearing, and turbulent motions, which can lead to overestimated outflow velocities \citep{concas2022being, 2023arXiv231006614X}. Second, the distinction between narrow and broad components is not standardized, and low-velocity outflowing gas can be misclassified as part of the narrow component. This misclassification may bias both the inferred velocity and mass of outflows upward \citep[e.g.,][]{2011ApJ...733..101G, schreiber2019kmos3d, swinbank2019energetics, 2019ApJ...873..122D, 2021MNRAS.503.5134A}.

When the total disc mass is fixed, outflow velocity shows little dependence on the initial gas fraction, as seen in models Z1F85D, Z1F70D, Z1F50D, and Z1F30D (Table~\ref{tab:parameters}). Although higher gas fractions generally boost SFR and feedback energy, this does not necessarily translate to faster outflows. In dense environments, a considerable fraction of the injected supernova energy is lost to rapid radiative cooling, which suppresses the growth and merging of hot bubbles \citep[e.g.,][]{2008IAUS..245...33C, fielding2018clustered}. Additionally, a thicker cold gas disk increases resistance to outflows, creating a ``blocking" effect that impedes their acceleration and vertical propagation. As a result, even under similar gravitational potentials, higher gas fractions do not always yield higher outflow velocities \citep[e.g.,][]{muratov2015gusty, 2020ApJ...900...61K}.

With a fixed stellar disk mass, decreasing the initial gas fraction results in increased outflow velocities for the warm phase but decreased velocities for the hot phase, as demonstrated by the simulations Z1F70D, Z1F50*S, and Z1F30*S. The reduced velocity of the hot gas in lower gas fraction simulations is likely due to diminished feedback energy. However, in these lower-density environments, the reduced resistance against cool and warm outflows and more efficient momentum coupling between hot and cool phases may enhance the acceleration of the cool and warm gas.

With the inclusion of gas return, a substantial fraction of gas stored in sink particles is recycled back into the ISM near the disk at temperatures of $\sim8000$ K. This increases the abundance of cool and cold gas in the outflow and enhances multiphase interactions. As a consequence, the hot phase experiences significant deceleration. As shown in Figure~\ref{fig:com}, the mean hot-phase velocities in Z1F70DG and Z1F50DG, measured at $t=30$ Myr and $r=3$ kpc, are substantially lower than those in Z1F70D and Z1F50D. A similar trend is reported in Figure 13 of \citet{2025ApJ...982...28L}. As we show in Section~\ref{subsec:time evolution and radial}, this deceleration of the hot phase occurs primarily at 
$r\gtrsim2$ kpc. Moreover, different gas phases evolve at different paces within the outflow, and the inclusion of gas return further amplifies this effect \citep{2025ApJ...982...28L}.

Meanwhile, the outflow velocities measured at $r=0.05R_{\mathrm{vir}}$ and $t=30$ Myr differ moderately between the `Z1' and `Z2' models. The mean velocities of the cold, cool, and warm phases are generally higher in the `Z1' runs than in the `Z2' runs. These differences primarily reflect variations in the initial conditions and the distinct evolutionary stages of the outflows in the two model sets.

\subsection{Mass Outflow Rate}\label{subsec:mass outflow rate}

\begin{figure*}[htb]
\begin{centering}
\hspace{-0.0cm}
\includegraphics[width=0.95\textwidth, clip]{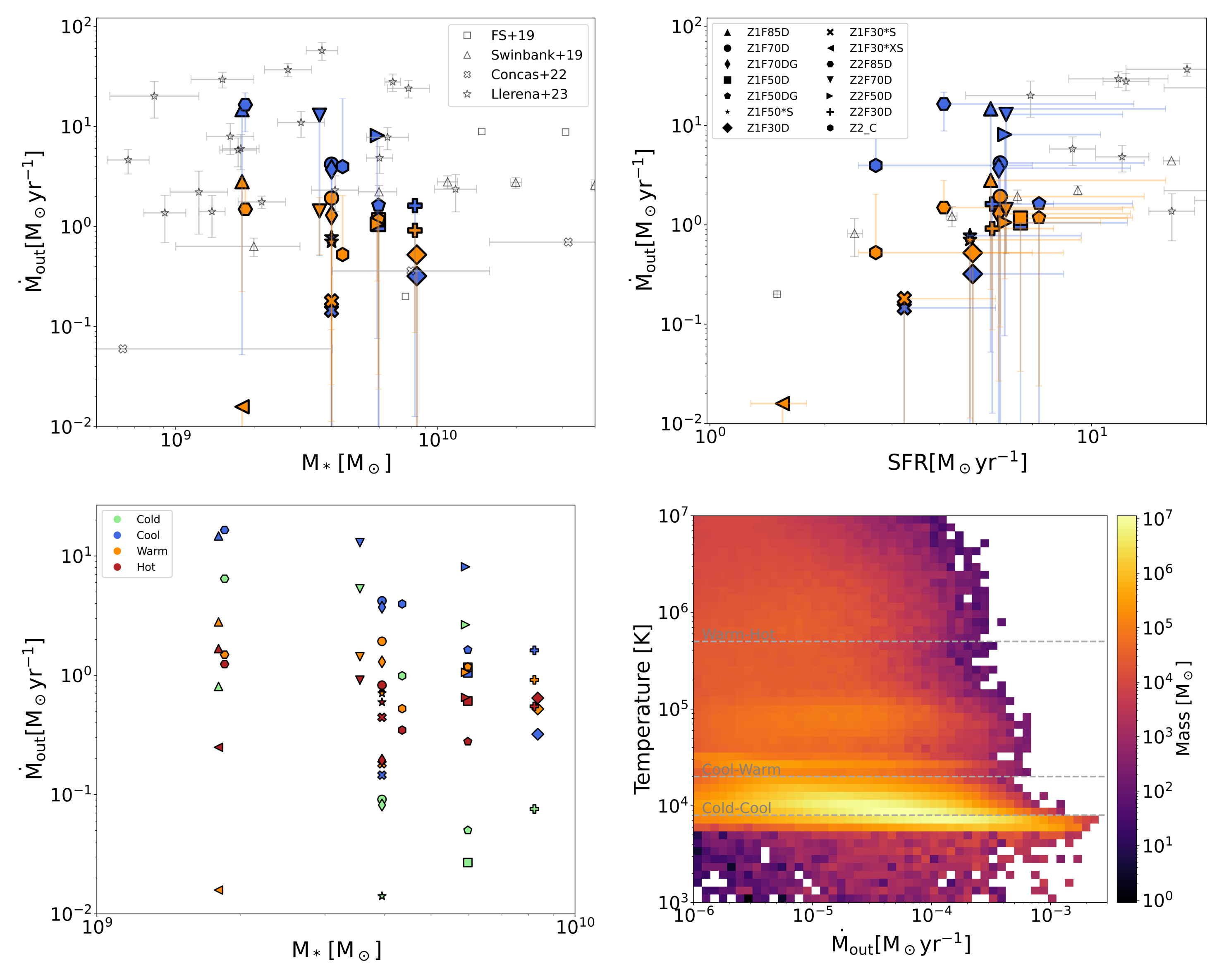}
\caption{Same as Figure~\ref{fig:com}, but showing mass outflow rates. Marker styles are identical to those in Figure~\ref{fig:com}. Gray open stars are low-mass star forming galaxies at $z \sim$ 3 from \citealt{llerena2023ionized}. Some data points in our simulations with mass outflow rates below $10^{-2}$ M$_\odot \,\rm{yr}^{-1}$ are not shown, as their contribution is negligible for the evolution of galaxies.}
\label{fig:comM}
\end{centering}
\end{figure*}

The mass outflow rate of a particular phase in each radial bin, $\dot M_{out}$, is calculated by 
\begin{equation}\dot{M}_{\mathrm{out}} = \sum_i \frac{\rho_i \, v_{\mathrm{rad},i} \, dx^3 }{\Delta r}
\end{equation}
where $\rho_i$ and $v_{\mathrm{rad},i}$ are the density and radial outflow velocity in a gas cell, $dx^3$ is cell volume, $\Delta r$ is the width of each radial bin.

At $t = 30$ Myr, the total mass outflow rates at $r=3$ kpc in our simulations range from 0.5 to 20 M$_\odot \,\rm{yr}^{-1}$. The cool (8000--$2 \times 10^4$ K) phase contributes up to approximately 17 M$_\odot \,\rm{yr}^{-1}$, while the cold (<8000 K), warm ($2 \times 10^4$--$5 \times 10^5$ K) and hot (>$5 \times 10^5$ K) phase ranges from 0.04 to 6 M$_\odot \,\rm{yr}^{-1}$, 0.02 to 3 M$_\odot \,\rm{yr}^{-1}$, and 0.2 to 2 M$_\odot \,\rm{yr}^{-1}$, respectively. The mass outflow rates of the cool and warm phases in our simulations are broadly consistent with observational estimates of warm ionized outflows, as shown in the top panel of Figure~\ref{fig:comM}. The bottom panel shows the distribution of gas cells in mass outflow rate–temperature space for Z1F70D at $0.05R_{\mathrm{vir}}$ and 30 Myr, color-coded by gas mass. Cool gas dominates the mass budget of the outflow, consistent with findings from some high-resolution idealized disk simulations \citep{2020ApJ...903L..34K}, though other studies report differing phase contributions \citep{nelson2019first, pandya2021characterizing}.

The relations between galactic wind mass outflow rates and galaxy stellar mass and SFR are important to galaxy evolution. However, because our simulations span a limited stellar-mass range ($10^9$–$10^{10}\,M_\odot$) and cover only a short evolutionary period (30 Myr), they cannot capture the long-term evolution of outflows. Accordingly, our analysis focuses on the underlying physical mechanisms and qualitative trends within this mass range. Quantitative scaling relations for mass loading factors are discussed in subsequent sections but should be interpreted with caution given these limitations.


Overall, our simulations show that the mass outflow rates of all four phases generally decrease with increasing stellar mass within $10^9$–$10^{10}\,M_\odot$, albeit with substantial scatter. Note that, the Z1F30*XS model exhibits both low SFR and low $\dot{M}_{\mathrm{out}}$ compared to typical observations, likely owing to its unusually low gas surface density. The robustness of this trend, particularly its slope, requires further investigation in the future due to the limited sample size and stellar mass range in our simulations. Nevertheless, it can be explained by several factors. In our default setup, all Z1' runs share the same total disk mass, while all Z2' runs also share a common disk mass, but with a different value. As a result, a higher initial stellar mass implies a lower gas fraction and gas mass, which is consistent with the general observational trend that more massive galaxies tend to have lower gas fractions \citep[e.g.,][]{2012MNRAS.426.1178N, 2012ApJ...760....6M, 2014A&A...562A..30S, 2015ApJ...800...20G, 2019ApJ...878...83W, 2021A&A...648A..25Z}. This leads to weaker starbursts. Furthermore, with the scale length and height held fixed, a higher stellar mass yields a deeper gravitational potential well, which confines outflows to smaller volumes and narrower opening angles, thereby reducing the overall mass outflow rate.

However, our default setup of galaxy models is idealized and does not incorporate full cosmological evolution. The galaxy models are constructed to probe outflow in isolate galaxies under specific “redshift-like” initial conditions, whereas real galaxies can span a wide range of total disk masses at a given redshift. For example, at fixed stellar disk mass, different gas fractions can produce different outcomes: higher gas fractions tend to result in stronger starbursts and higher outflow rates, as shown by simulations Z1F70D, Z1F50*S, and Z1F30*S. Similarly, when the gas fraction is held constant, a larger stellar mass also implies a larger initial gas mass, as seen in the comparison between Z1F50*S and Z1F50D. In these cases, stronger starbursts and higher outflow rates can still occur. 

Despite these complexities, the general observational trend that more massive galaxies have lower gas fractions suggests that the overall anti-correlation between mass outflow rate and stellar mass is very likely to persist when a broader sample is considered, although the slope may be flatter than our results to some extent. These results highlight that both stellar mass and gas mass play crucial roles in determining outflow properties. Therefore, caution is warranted when interpreting the slope of scaling relations based on limited simulation sample, which may not fully capture the diversity of real galaxies. In addition, the inclusion of gas return in the simulations has only a minor impact on the mass outflow rate.

On the observational side, the mass outflow rate of warm ionized gas, which generally corresponds to the cool and warm phases in our definition, does not show a clear correlation with stellar mass in some studies \citep[e.g.,][]{swinbank2019energetics, llerena2023ionized}, while others report a weak to moderate positive correlation \citep[e.g.,][]{schreiber2019kmos3d, concas2022being}. This apparent discrepancy is likely due to a combination of factors, including limited sample sizes, substantial variation in gas fraction among galaxies with similar stellar masses, inclusion of some extreme case and significant uncertainties in the measured outflow properties \citep[e.g.,][]{2011ApJ...733..101G, newman2012sins, 2020ARA&A..58..157T}.

Meanwhile, the mass outflow rates of the cool, warm, and hot phases measured at $t=30$ Myr in our simulations increase with SFR, where the SFR is averaged over $t=25$–30 Myr. This trend is in agreement with several observational studies \citep[e.g.,][]{schreiber2019kmos3d, swinbank2019energetics}. However, no such correlation is found in the analysis by \citet{llerena2023ionized}. Our simulations show that higher initial gas fractions lead to stronger starbursts and consequently more powerful outflows, whereas gas-poor models exhibit weaker feedback and lower mass outflow rates (see also Fig.~\ref{fig:R.png}). However, the SFR and the mass outflow rate measured at $r=0.05R_{\mathrm{vir}}$ are not synchronized: the SFR typically peaks at $t\sim3$ Myr, while the mass outflow rate peaks later, at $t\sim20$–30 Myr, with the exact timing varying across gas phases and simulations. This temporal offset complicates the interpretation of scaling relations between mass outflow rate and SFR.

\subsection{Mass Loading Factor}\label{subsec:mass loading factor}

\begin{figure*}[htb]
\begin{centering}
\hspace{-0.0cm}
\includegraphics[width=1.0\textwidth, clip]{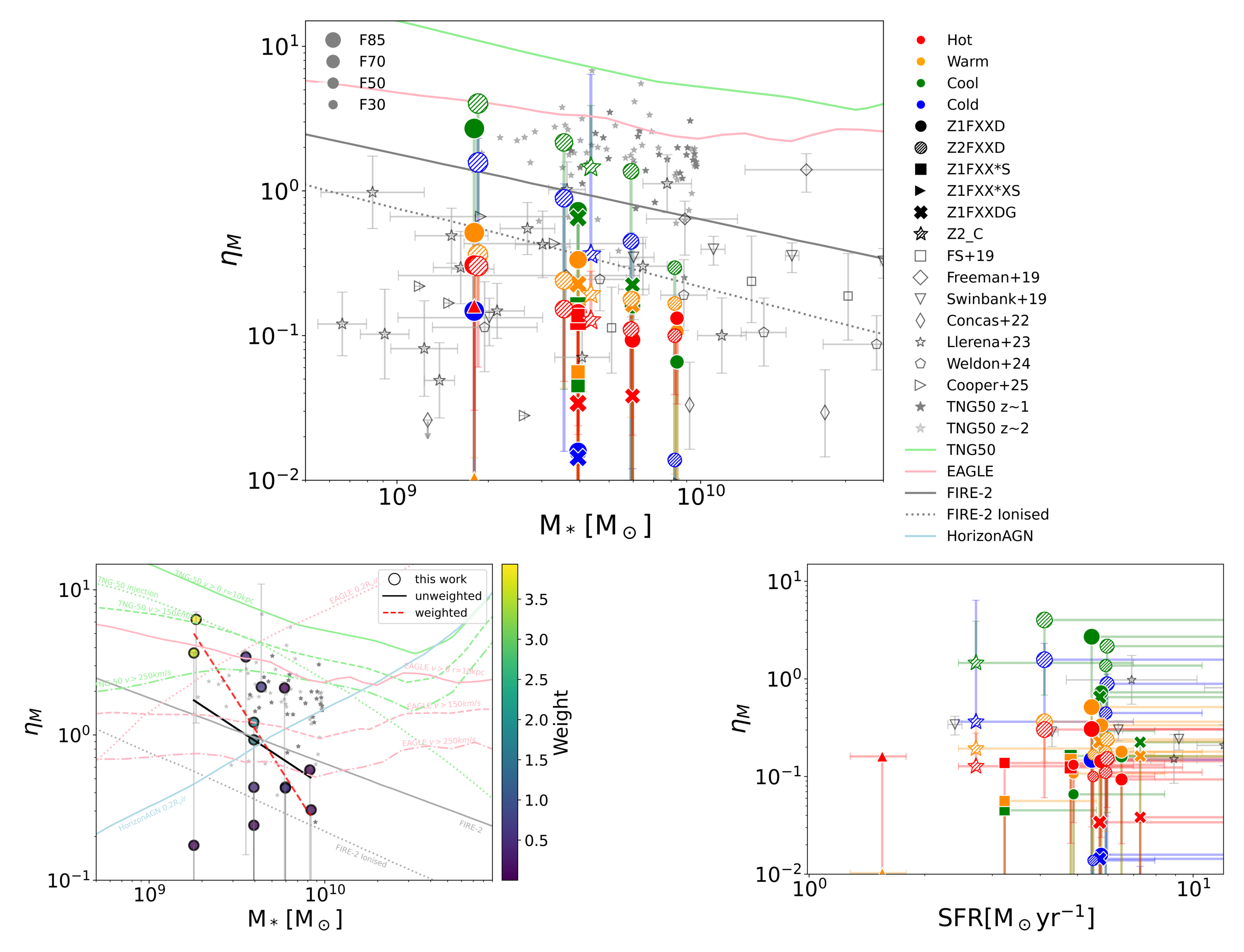}
\caption{Mass loading factor ($\eta_{\mathrm{M}}$) versus galaxy stellar mass ($M_*$) and star formation rate (SFR). Top: $\eta_{\mathrm{M}}$ of the cool (green) and warm (orange) phases versus $M_*$. Gray open symbols represent recent observations of warm ionized ($\sim 10^4-10^5$K) outflows in high redshift galaxies. For comparison, we also include predictions from other simulations: TNG50 (\citealt{nelson2019first}) in green line, EAGLE (\citealt{mitchell2020galactic}) in pink, and FIRE-2 in gray, with its ionized phase results shown as the gray dotted line (\citealt{pandya2021characterizing}). To facilitate a direct comparison of $\eta_{\mathrm{M}}$ at a consistent measurement radius, we calculate the $\eta_{\mathrm{M}}$ in TNG50 that match our samples in both galaxy properties and outflow measurement radius, which are shown as dark gray stars for $z \sim$ 1 and light gray stars for $z \sim$ 2. Bottom left: Total mass-loading factors from our simulations compared with those from other simulation studies, including Horizon-AGN \citep{2017MNRAS.467.4739K} (blue line). The black solid line shows an unweighted fit to our simulation results, while the red dashed line shows the weighted fit, with weights indicated by color (see text for details of the weighting scheme)} Bottom right: Mass loading factor as a function of SFR.
\label{fig:mload_com}
\end{centering}
\end{figure*}

We estimate the mass loading factor ($\eta_{\mathrm{M}} = \dot{M}_{\mathrm{out}}/\mathrm{SFR}$) separately for the cold, cool, warm, and hot phases, as well as for the total outflow across all phases in our simulations, also at 0.05 $R_{\mathrm{vir}}$ due to reasons given in the last subsection. To compute $\eta_{\mathrm{M}}$ at a given time $t$, we use the SFR averaged over the preceding 5 Myr, i.e., from $t - 5$ Myr to $t$. We first evaluate $\eta_{\mathrm{M}}$ at $t = 30$ Myr and then track its evolution back to $t = 18$ Myr. Figure~\ref{fig:mload_com} presents the resulting $\eta_\mathrm{M}$ values, along with their distributions as functions of stellar mass ($M_*$) and SFR. In the upper panel, solid markers in blue, green, orange, and hot represent $\eta_{\mathrm{M}}$ for the cold, cool, warm, and hot phases, respectively. The error bars indicate the maximum and minimum values during the $t =$ 18--30 Myr interval. 

In our simulations the total mass loading factor at 0.05 $R_{\mathrm{vir}}$ and $t = 30$ Myr spans nearly two orders of magnitude, ranging from $\sim 0.24$ to 6.26. The mass loading factor of cool phase, $\eta_\mathrm{M,c}$, less than 4.05, while the mass loading factor of warm phase, $\eta_\mathrm{M,w}$, lies in the range of 0.01 to 0.51. {Note that, the Z1F30*XS model adopts an extreme initial setup, i.e., a less massive, lower-density disk, leading to extremely low mass loading factors in the cool and warm phases ($\lesssim 0.01$).  At a stellar mass of $M_* = 10^{9.5}\,M_{\odot}$, the corresponding average values for the total, cool, and warm phases in our simulations are approximately 1.2, 0.75, and 0.25, respectively. The cool phase consistently exhibits a higher mass loading factor than the warm phase, which is expected given that the cool component dominates the outflow mass budget. Our estimated mass loading factors for the cool and warm phases are broadly consistent with recent observational studies of warm ionized outflows ($T_e \sim 10^4$--$10^5$ K) at cosmic noon, which report values in the range of $\sim 0.01$--3.0 for galaxies with stellar masses between $10^{8.5}$ and $10^{10.5} \,\rm{M_\odot}$ \citep{swinbank2019energetics, schreiber2019kmos3d, 2019ApJ...873..122D, 2021MNRAS.503.5134A, 2024MNRAS.531.4560W}.

Most recent observational studies of outflows adopt models based on \citet{2011ApJ...733..101G} and \citet{newman2012sins}, assuming spherical or multi-conical geometries and estimating $\dot{M}_{\mathrm{out}}$ using an opening angle \citep[e.g.,][]{swinbank2019energetics, schreiber2019kmos3d, 2019ApJ...873..122D, 2021MNRAS.503.5134A, 2024MNRAS.531.4560W, cooper2025high}. High-redshift galaxies ($z \sim 2$) often exhibit wide outflow cones \citep{venturi2018magnum, swinbank2019energetics, 2023arXiv231006614X}. In our simulations, $\dot{M}_{\mathrm{out}}$ and $\eta_{\mathrm{M}}$ are computed over the entire domain, potentially encompassing larger solid angles than assumed in observations. To facilitate comparison, we also evaluate these properties using a half-opening angle of $50^\circ$, which yields cool/warm mass loading factors in closer agreement with observed values (see Appendix~\ref{app:multiangle}).

While the mass-loading factors of the cool and warm phases in our simulations are broadly consistent with recent observational estimates, the comparison is not strictly apple-to-apple. First, observational outflow properties are inferred from emission or absorption lines that are broadened or shifted relative to the systemic velocity, and depend on assumptions about outflow geometry, projection effects, and ionization corrections (see \citealt{llerena2023ionized}). In contrast, outflow properties in our simulations are measured directly. Second, simulations define mass outflow rates as mass fluxes through a specified shell or surface, whereas observations infer them from integrated line-of-sight motions, often at different radii or heights. Third, our simulations model a single, idealized starburst over a limited spatial and temporal range, and the galaxy models may differ in scale and morphology from observed systems. These differences in methodology, definitions, and physical setup introduce systematic uncertainties that should be kept in mind when comparing simulation results with observations.

In our simulations (excluding the Z1F30XS model), the mass loading factors of both the cool and warm phases decrease with increasing stellar mass. Owing to its extreme initial conditions, the Z1F30XS model produces negligibly weak cool and warm outflows and is therefore excluded from this analysis. This trend arises because, across our simulations, the mass outflow rate declines more steeply than the SFR as the initial $M_*$ increases. In our default setup, galaxies with higher stellar mass have smaller gas reservoirs, weaker starbursts, and deeper potential wells, all of which contribute to more confined outflows with narrower opening angles and lower mass outflow rates. These results highlight the importance of caution when interpreting scaling relations between outflow properties and stellar mass, particularly when sample sizes are limited. 

The anti-correlation between the mass-loading factor $\eta_{\mathrm{M}}$ and $M_*$ for the cool and warm phases in our simulations is consistent with analytical expectations and previous simulations, which find that feedback is more efficient at driving outflows in low-mass galaxies than in their higher-mass counterparts \citep{murray2005maximum, muratov2015gusty}. However, observational results remain mixed. For instance, \citet{swinbank2019energetics} reports $\eta_{\mathrm{M}} \sim 0.1$--0.4 with a positive trend, $\eta_{\mathrm{M}} \propto M_*^{0.26 \pm 0.07}$, at $z \sim 1$. Similarly, \citet{2019ApJ...873..102F} finds $\eta_{\mathrm{M}} \sim 0.26$--1.4 for 127 star-forming galaxies and suggests a possible positive correlation with $M_*$. 

Other studies report negative or flat trends. \citet{concas2022being} using a revised estimation method, find $\eta_{\mathrm{M}} \sim 0.03$--0.08 for $M_* < 10^{10.8}\, \rm{M}_\odot$, indicating a negative correlation. \citet{llerena2023ionized} measures $\eta_{\mathrm{M}} \sim 0.05$--3.26 and observes a decline above $M_* \sim 10^{10}\, \rm{M}_\odot$. Likewise, \citet{2024MNRAS.531.4560W} reports $\eta_{\mathrm{M}} \sim 0.02$--1.44 with $\eta_{\mathrm{M}} \propto M_*^{-0.45}$, although their stacked spectra suggest a nearly constant trend. \citet{cooper2025high}, using JWST data, observes that $\eta_{\mathrm{M}}$ increases with decreasing stellar mass for $M_* < 10^9\, \rm{M}_\odot$. Finally, \citet{schreiber2019kmos3d} finds relatively low and roughly constant values of $\eta_{\mathrm{M}} \sim 0.1$--0.2 for stellar feedback-driven outflows. 

It is important to note that $\eta_{\mathrm{M}}$ can vary significantly at a given stellar mass in observations. This scatter likely arises from several factors, including intrinsic variations in gas fraction and outflow efficiency at different evolutionary stages, both highlighted in our simulations, as well as limited sample sizes and uncertainties in the estimation methods. Observational estimates of $\eta_{\mathrm{M}}$ are particularly sensitive to assumptions about electron density ($n_e$), electron temperature ($T_e$), and outflow radius ($R_{\mathrm{out}}$) \citep[e.g.,][]{2011ApJ...733..101G, 2019ApJ...873..122D, 2020MNRAS.491.1427S, 2023arXiv231006614X, 2024MNRAS.531.4560W}. For instance, increasing $n_e$ from $\sim 50$ cm$^{-3}$ \citep{2019ApJ...873..102F} to 380 cm$^{-3}$ can lower $\eta_{\mathrm{M}}$ from $\sim 0.64$--1.4 to $\sim 0.08$--0.2 \citep{2019ApJ...873..122D}. Given these methodological differences and the small observational samples, the $\eta_{\mathrm{M}}$--$M_*$ relation in observations remains uncertain and calls for confirmation using larger datasets and more robust, standardized measurement techniques. For a more detailed discussion, see Section~\ref{Comparison Observations}.

In the upper panel of Figure~\ref{fig:mload_com}, we also present the mass loading factors from one cosmological zoom-in simulation: FIRE-2 \citep{pandya2021characterizing}, and two cosmological volume simulation suites: TNG50 \citep{nelson2019first}, and EAGLE \citep{mitchell2020galactic}. All show a similar overall trend that $\eta_{\mathrm{M}}$ decreases with increasing stellar mass. Our results are broadly consistent with FIRE-2, with $\eta_{\mathrm{M,w}}$ in our simulations being lower by approximately 0.21 dex than the ionized gas values reported in FIRE-2 over the stellar mass range $1.8 \times 10^9$ to $8.4 \times 10^9$ M$_\odot$. However, methodological differences exist in how outflows are identified and measured among the simulations, which we discuss in more
detail in Section \ref{Comparison Simulations}.

While observational studies typically measure the mass loading factor of individual gas phases, most often warm ionized or cold neutral gas \citep[e.g.,][]{swinbank2019energetics, 2020A&ARv..28....2V, concas2022being, 2024MNRAS.531.4560W}, simulation studies usually present the total, multi-phase $\eta_{\mathrm{M}}$. Ignoring this difference can artificially amplify apparent discrepancies. For example, the warm-phase $\eta_{\mathrm{M,w}}$ in our simulations is comparable to observational values and is $\sim$0.21 dex lower than the ionized gas values in FIRE-2. However, it is significantly below the total $\eta_{\mathrm{M}}$ reported in FIRE-2, EAGLE, and TNG50 by $\sim$0.64, 1.19, and 1.56 dex, respectively, across the same stellar mass range. This highlights that part of the simulation–observation discrepancy arises from differences in statistical definitions.

In the bottom-left panel of Figure~\ref{fig:mload_com}, we present the total mass loading factor, $\eta_{\mathrm{M}}$, versus stellar mass and compare it with FIRE-2, EAGLE, and TNG50. In our simulations, total $\eta_{\mathrm{M}}$ ranges from 0.24 to 6.26 and generally decreases with $M_\ast$. An unweighted linear fitting of our results, indicated by the solid black line, are broadly consistent with FIRE-2 in the stellar mass range $1.8 \times 10^9$--$8.4 \times 10^9$ M$_\odot$, showing an average offset of $\sim$0.05--0.15 dex. This agreement may be partly coincidental, as our simulations have limited mass range, short durations, and small sample size, and differences in outflow definitions and measurement radii can underestimate true discrepancies (see Section~\ref{sec:discuss}). By contrast, EAGLE and TNG50 report systematically higher $\eta_{\mathrm{M}}$ than our simulations by $\sim$0.50 dex and 0.86 dex, respectively, for galaxies of comparable mass. Notably, HorizonAGN shows a contrasting positive correlation between $\eta_\mathrm{M}$ and $M_*$ (measured at 0.2 $R_{\mathrm{vir}}$), opposite to trends in the other simulation suites.

The apparent discrepancies in the total $\eta_{\mathrm{M}}$ between our high-resolution idealized simulations and previous cosmological simulations likely arise from a combination of differences in outflow measurement methodologies, baryonic physics implementations, and simulation timescale (see Section~\ref{Comparison Simulations} for more discussion). In particular, the radius at which outflow properties are measured can play a significant role. In our analysis, we adopt a radius of 0.05 $R_{\mathrm{vir}}$ (approximately 2.5--3 kpc), significantly smaller than the fixed radius of 10 kpc used in EAGLE and TNG50 studies. To assess the impact of this choice, we analyzed a subsample of TNG50 galaxies at $z=1$ and 2, selected to match our sample in key properties: stellar mass $M_* = 10^9$--$10^{10}$ M$_\odot$, half-mass radii of 1.0--1.6 kpc, SFR > 2 M$_\odot \,\rm{yr}^{-1}$, circularity parameter $\epsilon > 0.2$, and flatness < 0.7.

For these galaxies, we measured the outflow rate $\dot{M}_{\mathrm{out}}$ and total mass loading factor $\eta_\mathrm{M}$ at 0.05 $R_{\mathrm{vir}}$, considering gas with positive radial velocity ($V_{\mathrm{rad}} > 0$). The resulting total $\eta_\mathrm{M}$ values range from 0.3 to 7, compared to 0.25–6.26 in our simulations. The corresponding $\eta_\mathrm{M}$--$M_*$ distribution, shown as dark gray stars in Figure~\ref{fig:mload_com}, exhibit better agreement with our simulations. The average $\eta_\mathrm{M}$ at this smaller radius exceeds our values by 0.28 dex---roughly one-third of the offset observed when measured at radius of 10 kpc from the original TNG50 analysis. The difference of 0.28 dex left should be a combined effect of the differences in factors such as implementation of baryon physics, sample size, and simulation time duration.

\begin{figure*}[htb]
\begin{centering}
\hspace{-0.0cm}
\includegraphics[width=1.0\textwidth, clip]{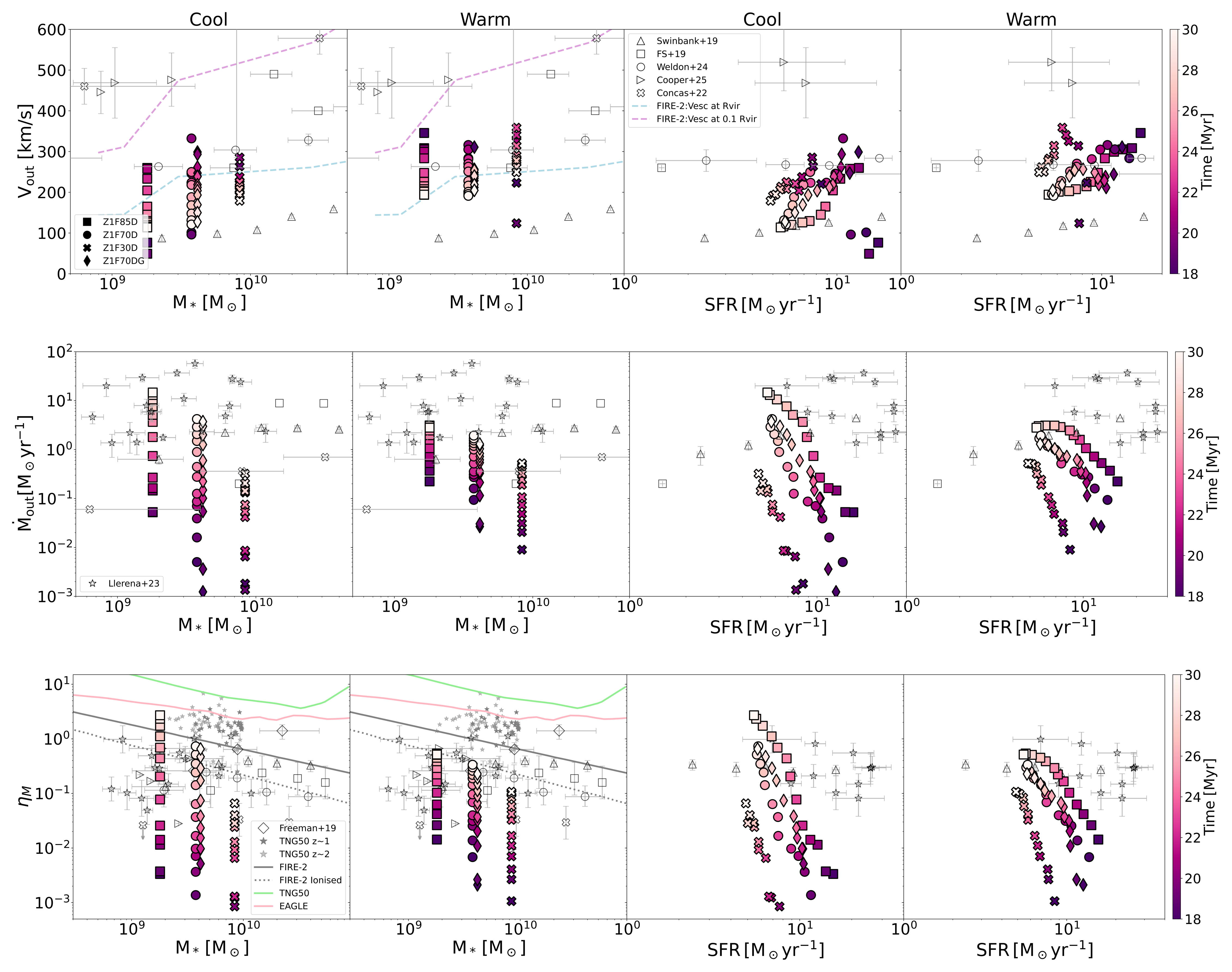}
\caption{The time evolution of outflow velocity (top), mass outflow rate (middle), and mass loading factor (bottom) from 18 to 30 Myr in four simulations (Z1F85D, Z1F70D, Z1F30D, and Z1F70DG). From left to right, the four columns show the properties of cool and warm phases versus stellar mass and SFR, respectively. Brighter colors indicate data more closer to t = 30 Myr.}
\label{fig:Vout_time}
\end{centering}
\end{figure*}

The scaling relation between the mass loading factor, $\eta_{\mathrm{M}}$, and stellar mass, $M_*$, is a key property of galactic outflows, as it reflects the underlying wind-driving mechanisms and strongly influences galaxy evolution. Owing to the limited sample size, narrow stellar-mass range, and short simulation duration, our data do not allow us to derive a robust $\eta_{\mathrm{M}}$–$M_*$ scaling relation. Nevertheless, as an exploratory exercise, we apply two fitting approaches to our results: an unweighted linear fit and a weighted linear fit, shown as the black solid and red dashed lines, respectively, in the bottom-left panel of Figure~\ref{fig:mload_com}.

The weighting scheme incorporates two observational facts. First, one weighting factor reflects the galaxy stellar mass function (GSMF) of star-forming galaxies on the main sequence, accounting for the fact that low-mass galaxies are statistically more abundant in the universe \citep{2025MNRAS.541..463H}. Second, we incorporate the redshift-dependent molecular gas fraction–stellar mass relation, $f_{\mathrm{gas}}-M_\ast$, and its scatter from \citet{2018ApJ...853..179T}. We convert the molecular gas fraction to a total (molecular + atomic) gas fraction by adopting a representative ratio of $M_{\mathrm{H_2}}/M_{\mathrm{HI}} = 0.3$ \citep[e.g.,][]{2017MNRAS.467..115D, 2020Natur.586..369C, 2022ApJ...935L...5C}. An additional weighting factor is assigned based on each simulation’s deviation from the $f_{\mathrm{gas}}$–$M_*$ relation. These two factors are multiplied and normalized prior to fitting. 

We argue that this weighting scheme improves the representativeness of our limited sample. Although both the unweighted and weighted fits show a decreasing trend of $\eta_{\mathrm{M}}$ with increasing $M_\ast$, the weighted fit yields a substantially steeper slope, changing from $-0.8$ to $-1.9$. This sensitivity highlights that scaling relations inferred from small or incomplete samples should be interpreted with caution. 

The bottom-right panel shows $\eta_{\mathrm{M}}$ versus SFR at $t=30$ Myr. In our simulations, the mass loading factors of the cool and warm phases display a weak positive trend with SFR, accompanied by substantial scatter. Observational results for these phases are mixed: some studies find no clear correlation \citep[e.g.,][]{swinbank2019energetics}, others report positive trends \citep[e.g.,][]{llerena2023ionized}, while still others find negative correlations \citep[e.g.,][]{cooper2025high, 2020ApJ...903L..34K}. Larger samples are therefore required to robustly constrain this scaling relation.

\subsection{Time evolution and radial profile}\label{subsec:time evolution and radial}

\begin{figure*}[htb]
\begin{centering}
\hspace{-0.0cm}
\includegraphics[width=1.0\textwidth, clip]{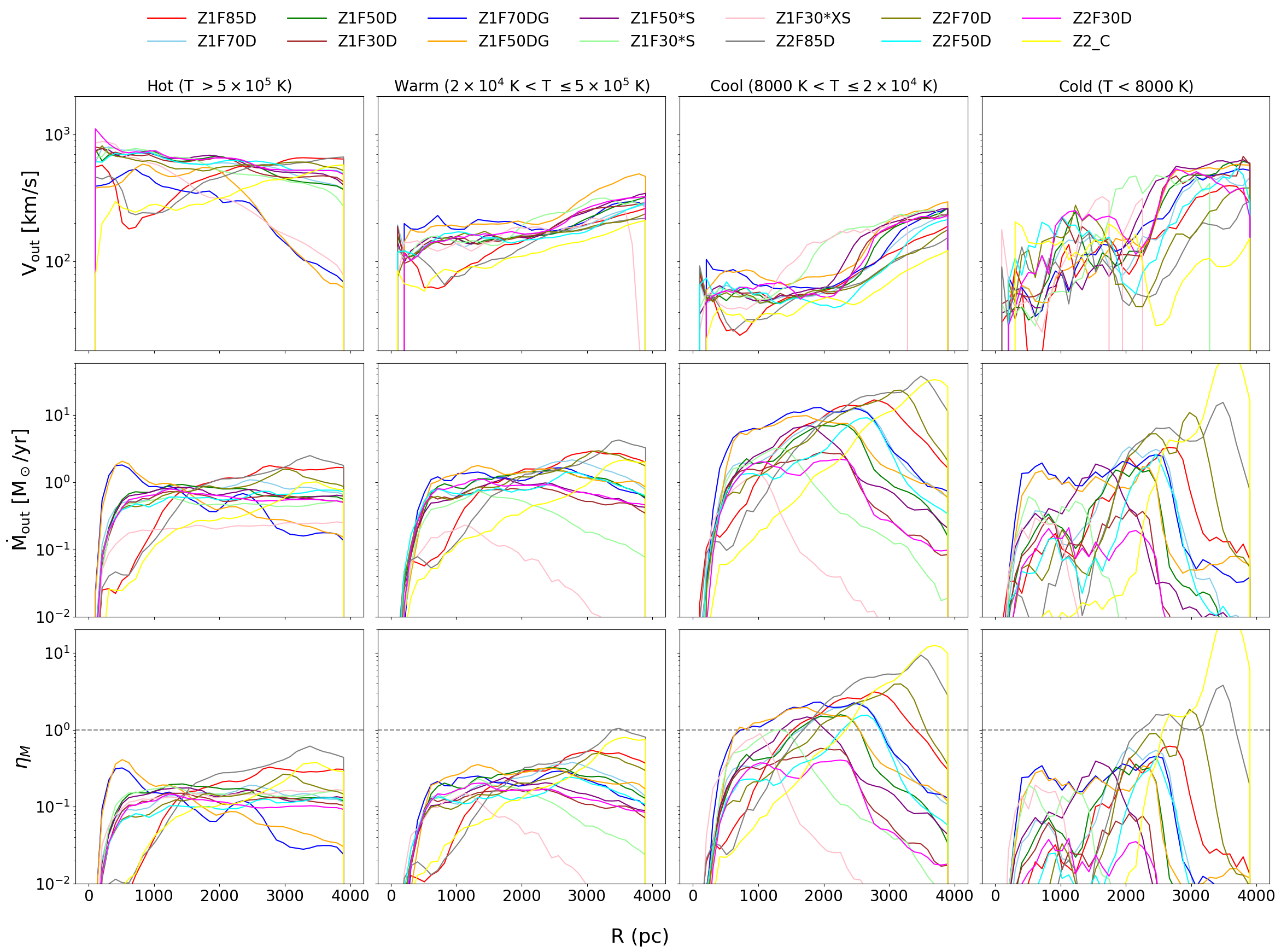}
\caption{Radial profiles of outflow velocity (top), mass outflow rate (middle), and mass loading factor (bottom) of the hot, warm, cool, and cold phases at $t = 30$ Myr. Solid lines in different colors represent various simulations, while the gray dashed line represents $\eta_{\mathrm{M}} = 1$. Due to differences in the timing of peak outflow activity across simulations and the fact that time evolution is not accounted for here, the values measured at 0.05$R_{\mathrm{vir}}$ exhibit noticeable variations.}
\label{fig:R.png}
\end{centering}
\end{figure*}

It is important to note that our simulations are halted at 30 Myr, which does not necessarily coincide with the peak of each galaxy’s outflow activity. As discussed earlier, outflow properties can vary substantially over time as the system evolves. To illustrate this temporal variation, we selected four representative simulations---Z1F85D, Z1F70D, Z1F30D, and Z1F70DG---and tracked their outflow velocities, mass outflow rates, and mass loading factors from $t = 18$ to 30 Myr. These results are presented in Figure~\ref{fig:Vout_time}, alongside relevant observational data.

As shown in the top panel of Figure~\ref{fig:Vout_time}, outflow velocities evolve significantly between $t = 18$ and 30 Myr, though the specific evolutionary paths vary among the four simulations. The cool outflow velocity, $V_{\mathrm{out,c}}$, measured at 3 kpc, shows a rise followed by a decline in all cases, but the timing of the peak differs across simulations. In contrast, the warm outflow velocity generally decreases over time, except in Z1F30D, where a different trend emerges. These phase-dependent and simulation-specific variations are consistent with the findings of \cite{2025ApJ...982...28L}, who showed that the time evolution of each outflow phase is influenced by several factors, including the initial galaxy setup, gas disk properties, and the implemented baryonic physics.

In all cases, the full range of cool and warm outflow velocities between $t = 18$ and 30 Myr overlaps well with most observational data---except for outflows at $z \gtrsim 2.5$ reported by \citealt{cooper2025high}. This agreement supports the validity of our multiphase outflow modeling in relatively low mass galaxies at cosmic noon. Furthermore, our results highlight the need for caution when interpreting scaling relations between outflow properties and global galaxy parameters such as stellar mass or SFR, as the timing of observational snapshots relative to the starburst phase remains inherently uncertain.

The middle panel of Figure~\ref{fig:Vout_time} shows that $\dot{M}_{\mathrm{out}}$ at a radius of 3 kpc increases with time, likely due to the growing volume and density of the outflow at this radius during the examined interval. The magnitude and rate of this increase vary across different gas phases and simulations, reflecting the diversity in the evolution of outflow driven by differing initial conditions and implemented physical processes. When plotted against SFR, the mass outflow rates of both the cool and warm phases trace a similar path---shifting from an initial high-SFR, low-outflow regime to a later low-SFR, high-outflow state. This behavior suggests a delayed response of the outflow to changes in star formation and feedback activity. Notably, the mass outflow rates in our simulations do not reach the range observed in current data until after $t\sim$20 Myr, implying that a certain timescale is required for outflows to accumulate and transport sufficient material beyond the disk to match observed intensities.

The bottom panel of Figure~\ref{fig:Vout_time} illustrates the time evolution of $\eta_{\mathrm{M}}$, which increases rapidly after $t = 18$ Myr, reflecting the cumulative effects of the starburst and the continued injection of supernova energy. By approximately $t = 23$ Myr, $\eta_{\mathrm{M}}$ in all simulations rises above the observational threshold of 0.01. Simulations with lower initial gas fractions exhibit both lower values of $\eta_{\mathrm{M}}$ and slower growth, primarily due to less intense starburst activity. The inclusion of the gas return process accelerates the increase in $\eta_{\mathrm{M}}$ between $t = 18$ and $t = 30$ Myr, although it results in only a modest change in the final $\eta_{\mathrm{M}}$ at the end of the simulation. 

In addition to strong time evolution, multiphase outflows also exhibit pronounced radial variations (e.g., \citealt{2023arXiv231006614X, 2025ApJ...982...28L}). Figure~\ref{fig:R.png} shows the radial profiles of outflow velocity, mass outflow rate, and mass loading factor for the hot, warm, cool, and cold gas phases across different simulations at $t = 30$ Myr. Consistent with the findings of \cite{2025ApJ...982...28L}, these properties can vary significantly with radius within a single simulation. However, in many simulations the mass outflow rate and mass loading factor of the cool and warm phases change moderately between $r=1.5$ and $3.0$ kpc. The radial trends also differ markedly among simulations, reflecting the sensitivity of outflow structure to initial conditions. These spatial and temporal offsets in the peak values of outflow properties can lead to apparent discrepancies when comparing results measured at fixed radii and epochs. 

Although we adopt a radius of 0.05 $R_{\mathrm{vir}}$ to enable direct comparison across simulations, intrinsic differences in initial gas conditions and feedback strengths still produce diverse outflow morphologies and propagation behaviors. In some simulations, the peaks of outflow velocity and mass outflow rate have not yet reached this radius by $t = 30$ Myr, while in others, these peaks have already moved beyond it. 

This challenge also extends to observations, where the time and radius where wind properties are measured can vary between studies, yet galactic winds are inherently time-dependent, unsteady, and multiphase phenomena (e.g., \citealt{2014ApJ...794..156R, 2019MNRAS.483.4586F, concas2022being}). Consequently, both studies based on simulations and observations must consider the full temporal and spatial evolution of galactic outflows. Measurements confined to a single radius and time snapshot, especially from small samples, may distort the wind’s dynamic behavior and cause systematic biases in inferred quantities and scaling relations. Nevertheless, the mass outflow rate and mass loading factor of cool, warm and hot phases change gently with radius within the range $1.5-3.0$ kpc. As a result, comparisons between our simulations and observations, typically measured at radii of $\sim$1–5 kpc, are likely only moderately affected by differences in the measurement radius. However, such comparisons may still be biased by differences in the evolutionary stage of the outflow.

\section{Discussions}
\label{sec:discuss}

\subsection{Comparison to observations}\label{Comparison Observations}

As shown in the previous section, the properties of cool and warm phase outflows in our simulations generally align with recent observations of ionized warm outflows at cosmic noon in terms of outflow velocity, mass outflow rate, and mass loading factor \citep{schreiber2019kmos3d, swinbank2019energetics, llerena2023ionized, 2023arXiv231006614X, 2024MNRAS.531.4560W}. However, we note that our simulations are limited in sample size and duration, and there are substantial differences in how wind properties are defined and measured between our simulations and the observational studies.

Moreover, our simulated outflow velocities are lower by approximately $50\%$ compared to the measurements reported by \citet{concas2022being}.
Additionally, our simulations differ from observations in the scaling relations between outflow and galaxy properties such as stellar mass and SFR. For example, we find  negative correlations between mass outflow rate, mass loading factor and stellar mass, whereas observational studies show significant variation in these trends. Both positive \citep{swinbank2019energetics} and negative \citep{concas2022being, llerena2023ionized, 2024MNRAS.531.4560W, cooper2025high, 2025MNRAS.tmp.1585P} correlations have been reported, highlighting the current lack of consensus.

Our investigation suggests that several factors contribute to the discrepancies in the scaling relations observed across different observational studies, as well as between observations and simulations (including ours). First, outflow properties are influenced by multiple interrelated galaxy parameters, such as stellar mass, gas fraction, and gas surface density. Second, outflows exhibit strong spatial and temporal evolution within a single starburst-driven event, whereas observations of individual galactic outflow typically capture only a brief snapshot of the outflow at a particular stage. Third, substantial differences exist in the definitions and measurement methods of wind properties across various studies.

Moreover, estimates of outflow properties in high-redshift galaxies often rely on specific models and assumptions (e.g., \citealt{2011ApJ...733..101G, newman2012sins}), making the derived quantities highly sensitive to factors such as emission line fitting process, assumed outflow geometry (spherical vs. biconical) and temperate, electron density, and the chosen measurement radius. For instance, \citet{llerena2023ionized} showed that increasing the assumed electron density from 380 cm$^{-3}$ to 560 cm$^{-3}$ reduces the inferred mass loading factor $\eta_\mathrm{M}$ by about 0.16 dex. Similarly, \citet{schreiber2019kmos3d} found that adopting $n_e = 380$ cm$^{-3}$ results in mass loading factors 3--10 times lower than their earlier estimates. 

Temperature assumptions also introduce significant uncertainty: \citet{2024MNRAS.531.4560W} reported that assuming a temperature of $1.5 \times 10^4$ K (as in \citealt{2020MNRAS.491.1427S}) increases the mass outflow rate by a factor of 1.5, while adopting a higher temperature of $2 \times 10^4$ K (as in \citealt{2011ApJ...733..101G}, for collisionally excited emission) decreases it by roughly a factor of two. Likewise, variations in the radius where outflow properties are measured---e.g., 3 kpc in \citet{2019ApJ...873..102F}, 10 kpc in \citet{swinbank2019energetics}, or the effective radius $R_e$ in \citet{2024MNRAS.531.4560W}---can significantly alter the derived values.

Taken together with the typically limited sample sizes in individual observational studies, these factors severely limit the robustness of the inferred scaling relations. As we demonstrated with our limited sample, altering the weighting scheme in the fitting process can significantly change the slope of relevant scaling relations. To derive robust scaling relations for galactic winds, both observations and simulations require larger, more complete samples with consistently measured and accurately characterized wind properties.

\subsection{Comparison to other Simulations}\label{Comparison Simulations}

In current cosmological simulations such as IllustrisTNG, EAGLE, SIMBA, and FIRE-2 (\citealt{2018MNRAS.473.4077P, 2015MNRAS.446..521S, 2019MNRAS.486.2827D, 2018MNRAS.480..800H}), the predicted properties of galactic outflows vary substantially. These discrepancies stem from differences in sub-grid baryonic physics as well as from the diverse methodologies employed to identify and measure outflow properties. In terms of the mass loading factor, $\eta_\mathrm{M}$, our results are broadly consistent with those from FIRE-2, differing by up to $\sim$ 0.2 dex across individual gas phases and in total, within the stellar mass range of $1.8 \times 10^9$ to $8.4 \times 10^9$ M$_\odot$. Both our simulations and FIRE-2 show a negative correlation between $\eta_\mathrm{M}$ and stellar mass. This agreement likely arises from the shared implementation of supernova feedback: both adopt a resolution-dependent hybrid thermal-kinetic model following \citet{kim2017three}. Nonetheless, the differences between our results and FIRE-2 may be somewhat underestimated, as they also reflect variations in how outflowing gas is defined and in the radial locations where outflow properties are measured.

In comparison, our total mass loading factors within the stellar mass range of $1.8 \times 10^9$ to $8.4 \times 10^9$ M$_\odot$ are lower than those reported in EAGLE (\citealt{mitchell2020galactic}) and TNG50 (\citealt{nelson2019first}) by approximately 0.50 dex and 0.86 dex, respectively, although the negative scaling of $\eta_\mathrm{M}$ with stellar mass is similarly present. It is important to note that varying definitions and measurement techniques across simulations likely contribute significantly to the apparent differences in outflow properties, potentially exaggerating the true physical discrepancies.

For example, we measure outflows at a radius of $\sim3$ kpc, while TNG50 and FIRE-2 focus on larger radii to assess the impact on the circumgalactic medium (CGM). Moreover, there is currently no standardized definition of ``outflow gas'' in either observations or simulations, making results highly sensitive to the adopted selection criteria. TNG50 identifies outflowing gas using several radial velocity thresholds ($V_{\mathrm{rad}} > 0$, 150, or 250 $\,\rm{km\,s}^{-1}$), whereas FIRE-2 applies a more stringent requirement: outflowing particles must have total Bernoulli energy at the measurement radius exceeding the gravitational potential at the target distance, ensuring they can escape into the ISM or halo. However, this criterion might systematically underestimate the mass loading factor, as it includes only gas that ultimately escapes the galaxy and excludes material that reaches large radii but remains gravitationally bound \citep{2024ApJ...966..129K}. In contrast, our analysis uses the more inclusive criterion $V_{\mathrm{rad}} > 0 \,\rm{km\,s}^{-1}$.

A more meaningful comparison requires harmonizing the methods used to define and measure outflows. For instance, if we apply our criterion ($V_{\mathrm{rad}} > 0$) to gas in TNG50 and measure at 0.05 $R_{\mathrm{vir}}$, instead of the 10 kpc radius used in \citet{nelson2019first}, the discrepancy in total mass loading factor $\eta_{\mathrm{M}}$ between our results and TNG50 is reduced from 0.86 dex to just 0.28 dex. Additionally, as shown in \citet{pandya2021characterizing}, the strict velocity and energy-based criteria of FIRE-2 exclude low-velocity outflow components. If a more relaxed velocity cut and measurements at 0.2--0.3 $R_{\mathrm{vir}}$ are adopted, the results resemble those of FIRE-1, which aligns more closely with TNG50 in both methodology and outcomes.

Therefore, we expect that applying our measurement approach to other simulations would yield total mass loading factors lower than those reported for FIRE-2, EAGLE, and TNG50 by approximately 0.2--0.4 dex. This indicates that a significant portion of the discrepancies in the total $\eta_{\mathrm{M}}$, both among different simulations and between simulations and observations, may be attributed to methodological differences. Establishing consistent definitions and measurement procedures would help reduce these inconsistencies and enable more meaningful comparisons across studies.

On the other hand, the cool phase gas ($8000 < T \leq 2 \times 10^4$ K) dominates the mass outflow rate in our simulations, consistent with findings for dwarf galaxies in FIRE-2. In contrast, TNG50 predicts that the majority of the outflowing mass resides in the hot phase. Notably, for galaxies with $M_* = 10^9$--$10^{10}$ M$_\odot$, the peak mass outflow rate in the hot phase exceeds that of the cool phase by an order of magnitude in TNG50. 

In addition to cosmological hydrodynamical simulations, supernova-driven galactic outflows have been extensively studied in recent high-resolution idealized simulations of isolated galaxies or gas disks (e.g., \citealt{2017MNRAS.470L..39F, 2020ApJ...903L..34K, schneider2020physical, schneider2024cgols, 2025MNRAS.539.1706V}). These studies generally produce multiphase outflows with broadly similar morphologies and global properties. They also find strong temporal and radial evolution during the first few tens of Myr, consistent with our results. Nevertheless, noticeable differences remain in the predicted wind properties across simulations. 

\citet{2017MNRAS.470L..39F} studied the development of multiphase outflows in isolated gas disks over 300 Myr at a resolution of $\Delta x = 3$ pc, sufficient to resolve the supernova cooling radius and inject feedback following the Sedov–Taylor solution (with $\sim$28\% kinetic and $\sim$72\% thermal energy). They reported wind evolution broadly consistent with the classic model of \citet{1985Natur.317...44C}: the time- and volume-weighted mean temperature (measured within a $45^\circ$ opening angle) decreases from $\sim10^5$ K to $\sim2\times10^4$ K over 1–6 kpc, while radial outflow velocities rise from $\sim$60 to $\sim$150 km s$^{-1}$. The resulting mass loading factors, $\eta_\mathrm{M}\sim0.5$–3.0, show only weak dependence on height above the disk.

In our simulations, the outflow evolution more closely resembles the model of \citet{2022ApJ...924...82F}, as discussed by \citet{2025ApJ...982...28L}. Nevertheless, the warm and cool phase velocities in most of our runs (Figures~\ref{fig:R.png} and \ref{fig:angle_2}) exhibit trends similar to those in \citet{2017MNRAS.470L..39F}, albeit with moderately higher velocities. Our total mass-loading factors span a comparable range, though they show a mild increase with radius. Differences likely arise because \citet{2017MNRAS.470L..39F} adopt a smaller gas disk and average outflow properties over much longer timescales than in our study. 

\citet{schneider2020physical} and \citet{schneider2024cgols} used high-resolution simulations to study multiphase outflows in M82-like galaxies driven by prescribed starburst and supernova distributions. With starburst strengths slightly higher than in our most active model (Z1F85D), they generate multiphase outflows broadly similar to our high gas fraction models. Their results highlight the critical role of mixing between cool and hot phases in shaping outflow properties, consistent with \cite{2024arXiv241209452W, 2025ApJ...982...28L}. They report moderately higher outflow velocities in the cool and warm phases, but somewhat lower mass outflow rates than in Z1F85D. The inclusion of clustered supernova feedback in our simulations likely enhances the mass outflow rate.

More recently, the QUOKKA project \citep{2025MNRAS.539.1706V} systematically explore how supernova-driven outflows depend on gas surface density ($\Sigma_{\rm gas}$), metallicity, and the SN injection scale height by high-resolution three-dimensional simulations. At fixed metallicity and injection height, higher $\Sigma_{\rm gas}$ lead to higher SFRs and stronger feedback, which efficiently heats the ISM and produces a volume-filling hot phase. These conditions generate faster and more stable outflows, but strongly suppress the cold ($T<2\times10^{4}\,$K) and warm ($2\times10^{4}<T<10^{6}\,$K) phases, resulting in lower mass loading factors ($\eta_{\mathrm{M}}$) despite higher energy loading factors ($\eta_{\mathrm{E}}$). In contrast, models with low $\Sigma_{\rm gas}$ exhibit low SFRs and inefficient heating, preventing the development of a hot phase; their outflows are instead dominated by cold and warm gas, yielding high $\eta_{\mathrm{M}}$ but extremely low $\eta_{\mathrm{E}}$. Similar trends have been reported in other parsec-scale simulations \citep{2020ApJ...903L..34K, 2020ApJ...900...61K}. 

The trends in QUOKKA partly differ from our results, which likely reflects variations in evolutionary stage, measurement methodology, star formation prescriptions and initial conditions. In QUOKKA, $\eta_{\mathrm{M}}$ is measured in a quasi–steady state by averaging over the central $1\times1,\mathrm{kpc}$ region for $t\gtrsim75$ Myr, when high $\Sigma_{\rm gas}$ systems are dominated by a volume-filling hot phase. In contrast, our simulations at $t=30$ Myr remain strongly time-dependent, with the shell at $0.05\,R_{\mathrm{vir}}$ still containing substantial cool and warm gas; higher initial gas fractions further enhance the cool-gas mass in the outflow at this stage. Additional differences arise because QUOKKA measures vertical velocities of all gas and adopts a fixed SN scale height, whereas our simulations employ a self-consistent SN distribution, which can systematically affect the inferred $\eta_{\mathrm{M}}$.

In short, persistent discrepancies in outflow properties, particularly velocities and phase-specific mass loading factors, across different simulations likely arise from a combination of differences in outflow definition and measurement methods, galaxy models and evolution history, simulation resolution, and stellar feedback injection and coupling schemes. Isolating the impact of each factor requires controlled code-comparison studies that apply identical initial conditions across different simulation frameworks (e.g., \citealt{2011MNRAS.418..960V, 2012MNRAS.423.1726S, 2014ApJS..210...14K, 2016ApJ...833..202K, 2021ApJ...917...64R, 2024ApJ...962...29S, 2024ApJ...977..233C}).

Further systematic comparisons of star formation and feedback prescriptions are also essential. Our simulations explicitly resolve star formation through gas collapse and sink-particle formation, incorporating well-established methods from the literature (e.g., \citealt{1995MNRAS.277..362B, 2004ApJ...611..399K, Federrath2010, Gong2012, Howard2016}). In addition, we mitigate supernova overcooling by adopting the feedback prescriptions of \citet{kim2017three} and by accounting for enhanced momentum injection from clustered supernovae, thereby maintaining efficient feedback at our resolution.

\subsection{Caveats and future improvement}

It is important to note that, due to computational constraints, our simulations are limited to a relatively narrow stellar mass range of $\sim 10^9$--$10^{10}$ M$_\odot$, corresponding to the typical masses of star-forming galaxies at cosmic noon. Additionally, the number of simulated galaxies at each stellar mass is limited, and our galaxy models are based on a number of simplifying assumptions. Moreover, the simulations span only 30 Myr, limiting our ability to probe long-term evolution. Consequently, while we identify some trends between outflow properties and galaxy parameters within this mass range, their applicability at lower or higher masses, as well as the precise slopes of the scaling relations, remain uncertain and require further investigation.

For instance, in simulations with fixed stellar mass but varying initial gas fractions (e.g., Z1F70D, Z1F50*S, Z1F30*S), we find that the mass loading factor decreases with decreasing initial gas fraction. If this trend extends to more massive systems, it would tend to flatten the slope of the $\eta_{\mathrm{M}}$--$M_*$ relation. Consistent with our results, previous studies such as \citet{2017MNRAS.465.1682H} and \citet{pandya2021characterizing} suggested that higher gas fractions can improve the mass loading factor.

Furthermore, our current study omits several potentially important physical processes, such as cosmic ray feedback and Type Ia supernovae. These effects may significantly influence outflow properties. For example, \citet{2021MNRAS.501.4184H} demonstrated that cosmic rays can substantially alter the structure and dynamics of galactic winds. Additionally, our simulations cover relatively short timescales and small volumes, limiting our ability to trace the long-term evolution and eventual fate of outflow gas.

Future work should explore a broader parameter space, covering a wider range of galaxy masses, gas fractions, and disk morphologies, ideally within a cosmological context to capture long-term evolution and environmental effects on galactic winds. Incorporating additional physical processes, such as cosmic rays and AGN feedback, will be crucial for more realistic predictions. Furthermore, careful apple-to-apple comparisons between simulations and observations are needed to achieve a robust understanding of starburst-driven galactic winds.

\section{Conclusions}
\label{sec:conclusions}

In this study, we performed a suite of three-dimensional hydrodynamic simulations to investigate multiphase galactic winds in star-forming galaxies within the stellar mass range of $10^9$ to $10^{10}$ M$_\odot$ at cosmic noon ($z\sim$ 1--2). We modeled the star formation and outflow in 14 idealized isolate disc galaxies, adopting the framework of \cite{2024arXiv241209452W} and \cite{2025ApJ...982...28L}, which successfully reproduces the observed multiphase outflows in M82. Supernovae feedback is handled with resolution dependent thermal-kinetic compound module, while enhanced feedback effect due to clustered SN is also considered. Each simulation produces a starburst lasting $\sim$ 20--30 Myr that drives kpc-scale outflows. We analyzed the starburst and outflow properties, examined their correlations with host galaxy characteristics, and compared our results with recent observations and simulations. We explore how factors such as the initial stellar mass and gas fractions influence starburst and outflow evolution. Our results can be summarized as follows:

\begin{itemize}
\item Galactic outflows in our simulations have velocities of $V_{\mathrm{out}} \sim 50-1000\,\rm{km\,s}^{-1}$, mass outflow rates of $\dot{M}_{\mathrm{out}} \sim 0.3-20$ M$_\odot \,\rm{yr}^{-1}$, and total mass loading factors $\eta_\mathrm{M}$ ranging from $ \sim 0.24$ to 6.26. The properties of cool ($8000 < T \leq 2 \times 10^4$ K) and warm phases ($2\times10^4 < T \le 5\times10^5\,\rm{K}$) in the outflows are generally consistent with observations, with typical velocity 100--400 $\,\rm{km\,s}^{-1}$, mass outflow rate $0.01-17$ M$_\odot \,\rm{yr}^{-1}$, and mass loading factors ranging from 0.01 to 4. Cool phase dominates the mass outflow rate
in our simulations. At a stellar mass of $M_* = 10^{9.5}\,M_{\odot}$, the average values of mass loading factors for the total, cool, and warm phases are approximately 1.2, 0.75, and 0.25, respectively. 

\item Outflow velocity shows no clear correlation with stellar mass or SFR, whereas mass outflow rate and mass loading factor generally decrease with increasing stellar mass and increase with SFR. The total mass loading factor exhibits a negative trend with stellar mass, both in unweighted fitting and fitting weighted by stellar mass and gas fraction, broadly consistent with trends reported in previous simulations.

\item Outflow properties depend on both initial stellar mass and gas fraction. At fixed redshift and total disk mass, higher gas fractions produce stronger starbursts and more extended, wider-angle outflows. Outflows also evolve substantially in time and space, so comparisons at fixed times or radii with limited samples risk biasing conclusions about absolute values and scaling relations. This partly explains the diversity seen in observed trends between outflow properties and galaxy characteristics. Additionally, compact, gas-rich galaxy disks at higher redshifts tend to produce stronger outflows, driven by the enhanced intensity of starbursts.

\item Within the stellar mass range of $10^9$--$10^{10} M_\odot$, our total $\eta_{\mathrm{M}}$ is lower than those in EAGLE and TNG50 by 0.50 and 0.86 dex, respectively, but is slightly higher than that in FIRE-2 by 0.06 dex. However, differences in outflow measurement methods across simulations significantly affect these comparisons. When accounting for these methodological differences, discrepancies in the total $\eta_{\mathrm{M}}$ reduce to roughly 0.2--0.4 dex, suggesting that mismatches between observations and simulations reported in the literature could similarly be reduced. Nonetheless, the mass loading factors for a specific phase (e.g., the cool phase) can still differ notably between simulations.
\end{itemize}

Our results show that the starburst and outflow modeling framework developed for M82 in \cite{2024arXiv241209452W} and \cite{2025ApJ...982...28L} can be effectively applied to star-forming galaxies at cosmic noon. While broader simulations spanning a wider range of galaxy masses, gas fractions, morphologies, and realistic cosmic evolution are needed to further validate our findings. This work highlights the importance of accounting for stellar and gas mass, as well as the temporal and spatial evolution of galactic winds, when interpreting relations between outflow and galaxies properties. Moreover, larger sample sizes, consistent definitions and measurement methods, and apple-to-apple comparisons are essential for robustly comparing outflow properties across simulations and observations, and for better understanding starburst-driven galactic winds. 

\begin{acknowledgments}
We are grateful to the anonymous reviewer for his/her  valuable comments and suggestions, which have helped improve the manuscript. This work is supported by the National Natural Science Foundation of China (NFSC) through grant 12595314, 12173102 and 11733010. The calculation carried out in this work was completed on the HPC facility of the School of Physics and Astronomy, Sun, Yat-Sen University.

\end{acknowledgments}

\begin{contribution}
WSZ conceived the initial research concept and was responsible for writing and submitting the manuscript. HC conducted the simulations,  analysis and also contributed to the initial draft. XFL and TRW contributed to the framework of simulations and also contributed to the analysis, AK and LLF contributed to both the manuscript and the development of the research concept.
\end{contribution}

\appendix
\section{comparison of outflows in the `Z1' and `Z2' models}\label{app:redshift}
\renewcommand{\thefigure}{A\arabic{figure}}
\setcounter{figure}{0}

\begin{figure*}[htb]
\begin{centering}
\hspace{-0.0cm}
\includegraphics[width=0.70\textwidth, clip]{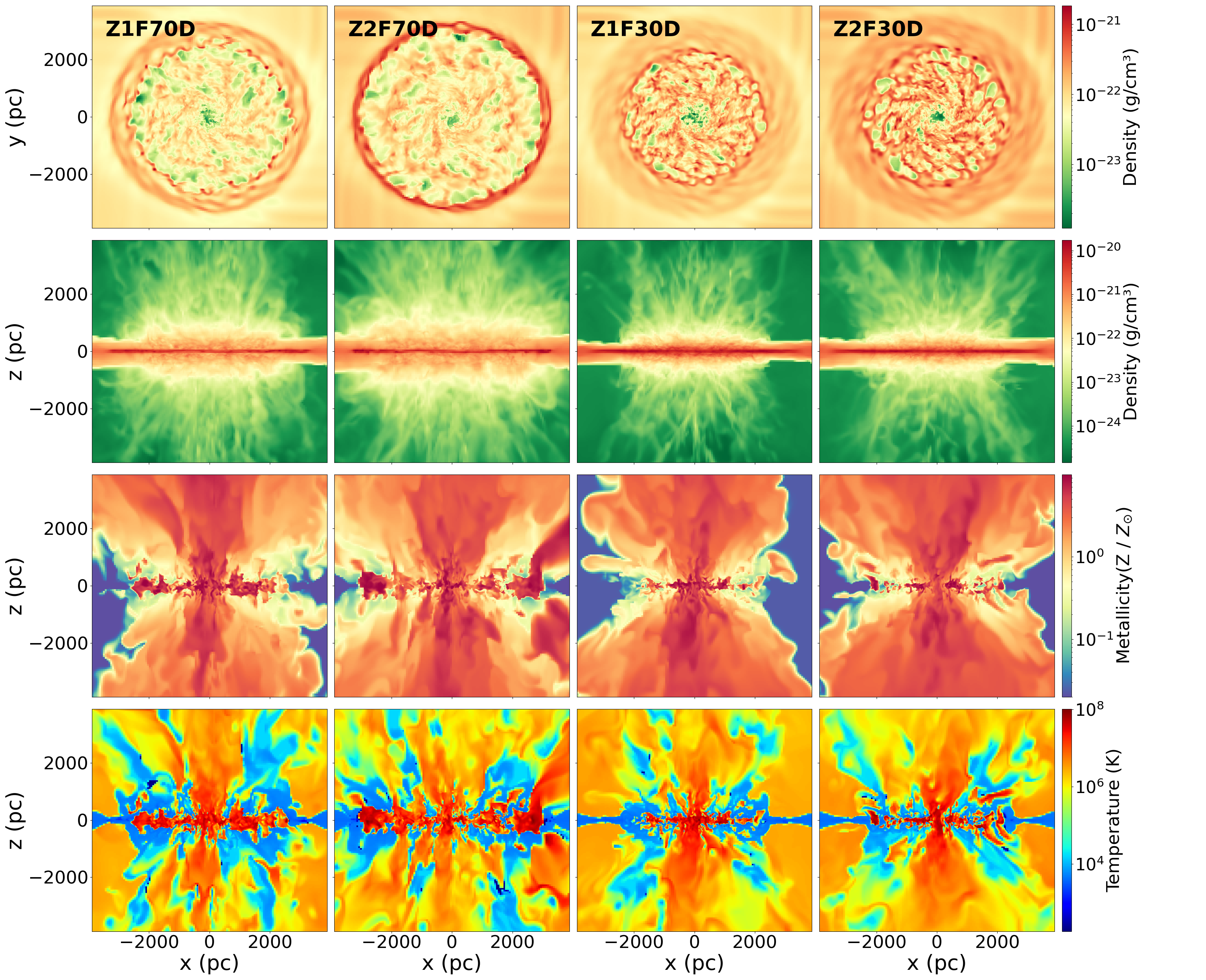}
\caption{Similar to Figure~\ref{fig:z1_530}, but showing a direct comparison between the `Z2' models (Z2F70D, Z2F30D) and their corresponding `Z1' counterparts (Z1F70D, Z1F30D). From top to bottom, each row of panels shows (1)face-on projected gas density, (2)edge-on projected gas density, (3) edge-on metal distribution, and (4)edge-on temperature distribution. }
\label{fig:z2_422}
\end{centering}
\end{figure*}

Figure~\ref{fig:z2_422} compares the outflow morphology of the `Z1' and `Z2' models. At fixed gas fraction, the Z2' models produce slightly more extended outflows than their Z1' counterparts, primarily due to differences in the initial conditions: the `Z2' models have more compact disks and higher gas column densities. As in the `Z1' models, lower gas fractions in the `Z2' models lead to more confined outflows. 

\renewcommand{\thefigure}{B\arabic{figure}} 
\setcounter{figure}{0} 
\setcounter{table}{0}                     
\renewcommand{\thetable}{B\arabic{table}} 
\section{impact of adopted opening angle}\label{app:multiangle}

\begin{figure*}[htb]
\begin{centering}
\hspace{-0.0cm}
\includegraphics[width=0.80\textwidth, clip]{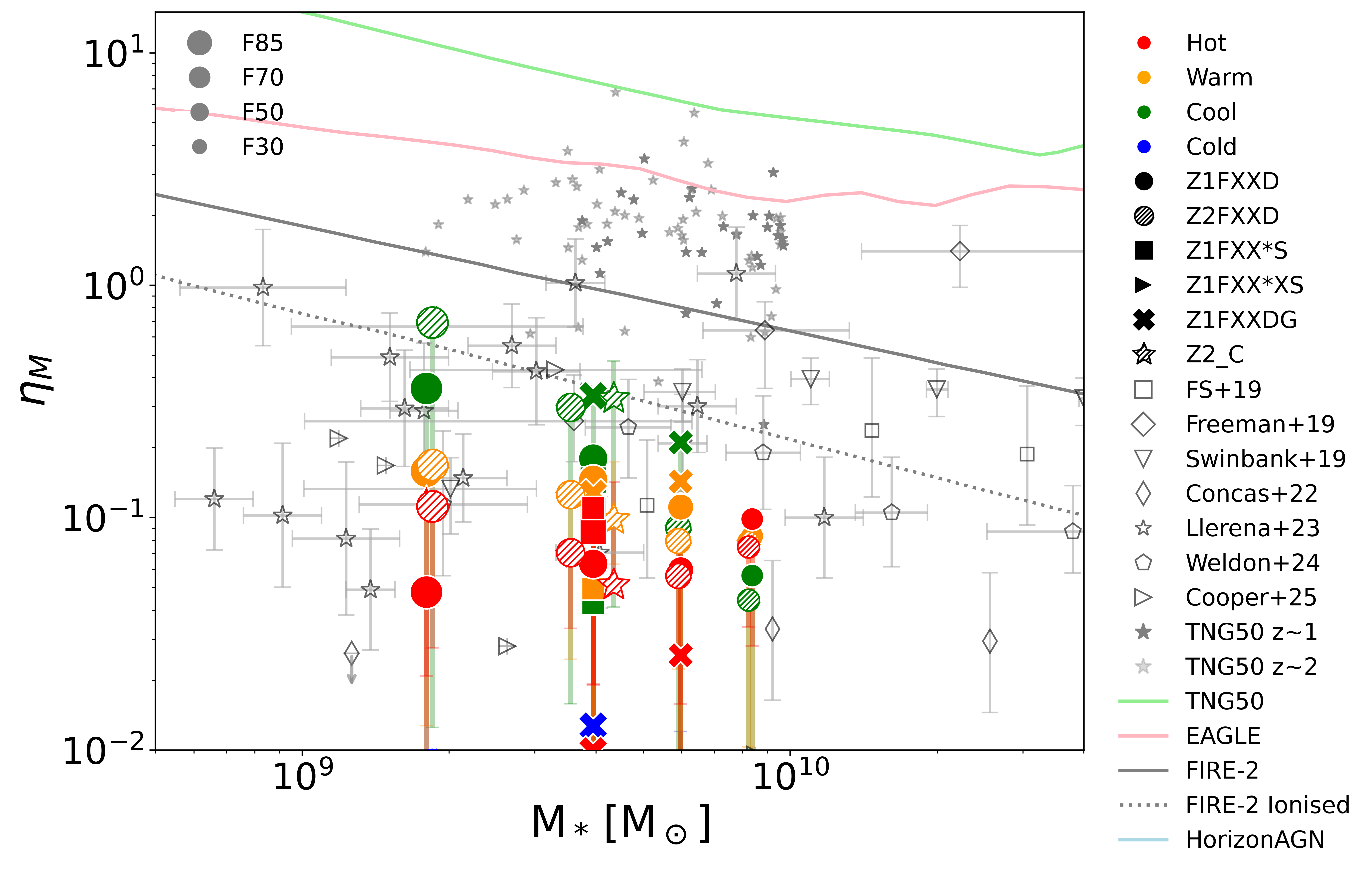}
\caption{Similar to Figure \ref{fig:mload_com} but calculated for a half-opening angle of $50^\circ$.}
\label{fig:Mload_M_obs}
\end{centering}
\end{figure*}

\begin{figure*}[htb]
\begin{centering}
\hspace{-0.0cm}
\includegraphics[width=0.75\textwidth, clip]{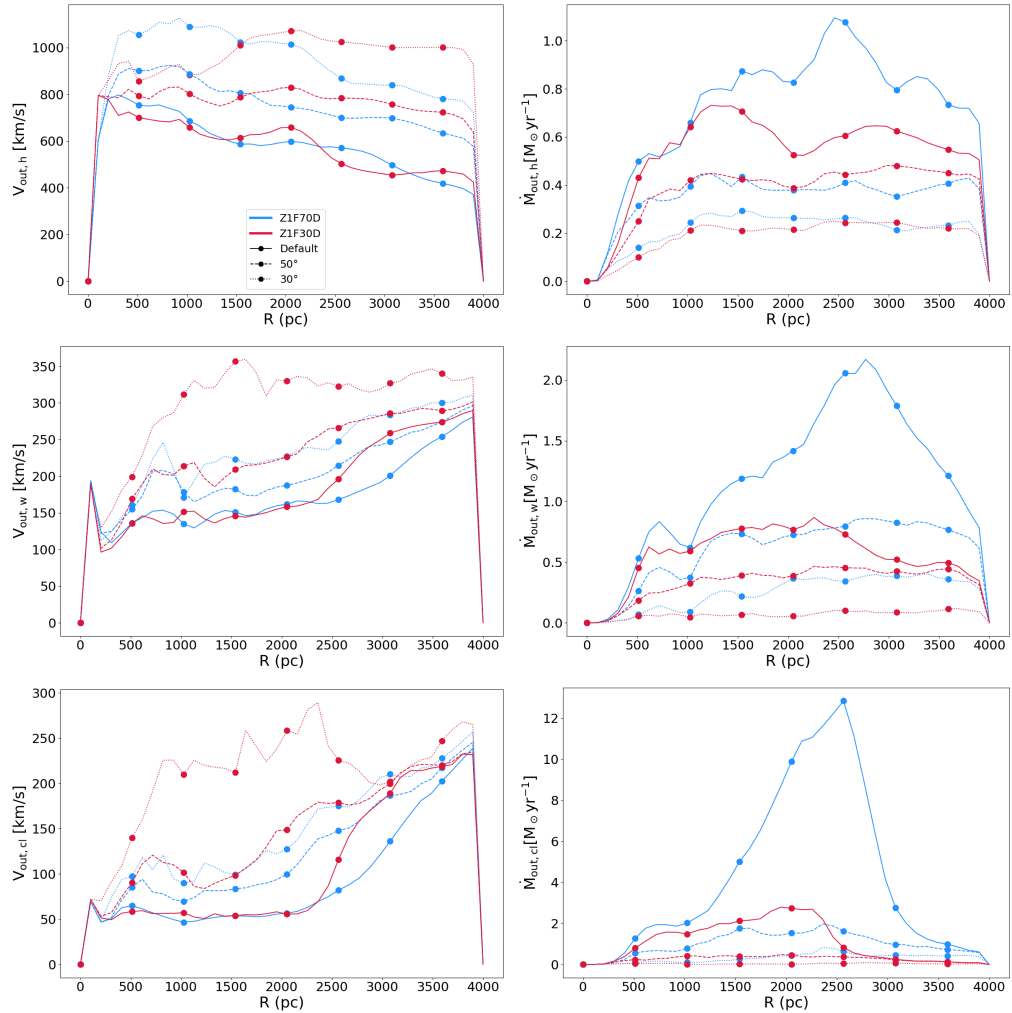}
\caption{Radial profiles of outflow velocity (left) and mass outflow rate (right) for the hot (top), warm (middle), and cool (bottom) phases in Z1F70D and Z1F30D, evaluated at the default angle ($90^\circ$) and half-opening angles of $50^\circ$ and $30^\circ$.}
\label{fig:angle_2}
\end{centering}
\end{figure*}

In this appendix, we present a quantitative analysis of how the adopted opening angle influences our results. Figure~\ref{fig:Mload_M_obs} shows the mass loading factor with a half-opening angle of $50^\circ$, instead of $90^\circ$ (the default angle, adopted in the main text). Furthermore, we select two representative simulations, Z1F70D and Z1F30D, to examine the impact of the adopted opening angle on the measured properties of outflow. We compare the radial profiles of $V_{\mathrm{out}}$ and $\dot{M}_{\mathrm{out}}$ under three different half-opening angles that nearly $90^\circ$, $50^\circ$ and $30^\circ$, as shown in Figure~\ref{fig:angle_2}. 

Figure~\ref{fig:Mload_M_obs} demonstrates that imposing a half-opening angle of $50^\circ$ leads to a significant reduction in the mass loading factors of the cool, and warm phases, with values reduced by a factor nearly 1--2. Figure~\ref{fig:angle_2} suggests that the variation in the adopted opening angle has a notable impact on the measured outflow velocity and mass outflow rate. Moreover, the variation of half-opening angle leads to opposite effects on outflow velocity and mass outflow rate. As the opening angle decreases, the mean outflow velocity increases, while the mass outflow rate declines. This is likely because narrower angles preferentially exclude low velocity gas located near the outflow edges. The associated decrease in mass outflow rate suggests that a significant fraction of the mass contribution arises from these excluded, low velocity components, particularly in the cool phase.

\vspace{15mm}




\bibliographystyle{aasjournalv7}
\bibliography{main}


\end{document}